\documentclass{IEEEojcsys}
\usepackage{amsmath,amsfonts}
\usepackage{algorithmic}
\usepackage{algorithm}
\usepackage{array}
\usepackage[caption=false,font=normalsize,labelfont=sf,textfont=sf]{subfig}
\usepackage{textcomp}
\usepackage{stfloats}
\usepackage{url}
\usepackage{verbatim}
\usepackage{graphicx}
\usepackage{amsthm}
\usepackage{xcolor}
\newtheorem{assumption}{Assumption}
\newtheorem{theorem}{Theorem}
\newtheorem{proposition}{Proposition}
\newtheorem{remark}{Remark}

\usepackage{bm}
\usepackage{cite}
\usepackage[colorlinks,urlcolor=blue,linkcolor=blue,citecolor=blue]{hyperref}
\makeatletter\renewcommand\p@subsection{\thesection-}\makeatother
\usepackage{booktabs}
\jvol{00}
\jnum{XX}
\paper{1234567}
\pubyear{2026}
\receiveddate{XX Month 2026}
\accepteddate{XX Month 2026}
\publisheddate{XX Month 2026}
\currentdate{XX Month 2026}
\doiinfo{OJCSYS.2026.Doi Number}

\begin{document}

\sptitle{Article Category}

\title{Policy-Conditioned Koopman Representations for Parameterized Closed-Loop Systems}

\author{KIRAN G. KRISHNAN\affilmark{1}}

\author{SHANKAR A. DEKA\affilmark{1} (Member, IEEE)}

\affil{Department of Electrical Engineering and Automation, Aalto University, 02150 Espoo, Finland}

\corresp{CORRESPONDING AUTHOR: Kiran G. Krishnan (e-mail: \href{mailto:kiran.gangadharannairkrishnan@aalto.fi}{kiran.gangadharannairkrishnan@aalto.fi})}

\authornote{This work was supported by NordForsk through the INFORM project under Grant 234623.}

\markboth{POLICY-CONDITIONED KOOPMAN REPRESENTATIONS FOR PARAMETERIZED CLOSED-LOOP SYSTEMS}{K. G. KRISHNAN AND S. A. DEKA}

\begin{abstract}
Predicting the response of a nonlinear system across different feedback controllers requires a model that captures how the closed-loop dynamics change with the controller parameters. In a finite-dimensional Koopman model, these changes may affect both the latent dynamics and the observable coordinates used to represent them. We propose a policy-conditioned Koopman representation framework to address this problem. It allows a single learned model to predict a family of closed-loop trajectories without retraining for each controller. We test the framework on six nonlinear prediction benchmarks, where the mean error of the model stays below a 15\% threshold on every system, while that of each baseline exceeds it on at least one. As an example application, the resulting surrogate supports one-shot feedback-parameter selection, in which an optimization at the initial state selects a controller that then operates with fixed parameters and ongoing state feedback. On nonlinear control benchmarks including MuJoCo, the selected controllers stabilize every system from the tested initial state with 1.4 to 21 times less online computation than nonlinear model predictive control, counting the selection itself. The results show that policy-dependent coordinates can support both prediction across closed-loop systems and controller selection without repeated online trajectory optimization.
\end{abstract}


\maketitle

\section{INTRODUCTION}
\label{sec:introduction}
The Koopman operator offers a route to linear prediction for
nonlinear systems. It acts on scalar observables by composing them with the dynamics, and this composition is linear on
the space of observables even when the underlying dynamics are
nonlinear~\cite{Koopman1931,mezic2005spectral,rowley2009spectral}. When a finite
collection of observables is close to invariant under the operator, nonlinear
state evolution can be approximated by applying a matrix to a latent vector and
decoding the result. Dynamic mode decomposition (DMD) makes this practical from
snapshot data, and its dictionary-based extension, extended DMD (EDMD), turns the problem into a
least-squares regression on lifted
coordinates~\cite{schmid2010dmd,Williams2015EDMD,williams2015kernel}. Since the
results depend heavily on the dictionary, the observables are now often learned from data, with encoder--decoder pairs trained so that their latent
dynamics are
linear~\cite{Lusch2018,takeishi2017learning,Otto2019LRAE,yeung2019learning,azencot20a_ConsistentKoopman}.
For control, the same construction yields a linear predictor that can be placed
inside a receding-horizon
problem~\cite{proctor2016dmdc,Korda2018KoopmanMPC,cibulka2025dictionaryfree,haseli2026kcf}.

\begin{figure}[!t]
\centering
\includegraphics[width=\columnwidth]{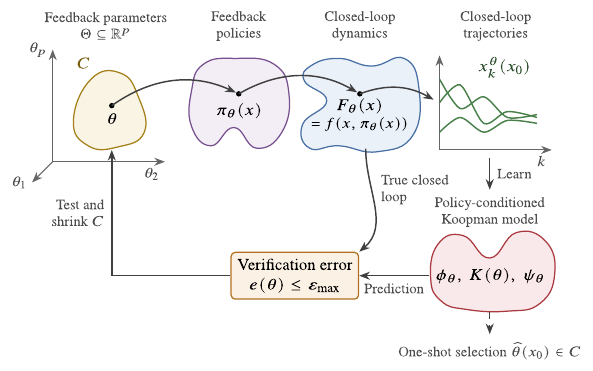}
\caption{Overview of the framework. Each feedback parameter $\theta\in\Theta$
defines a policy $\pi_\theta$ and a closed loop $F_\theta$, whose trajectories
train one policy-conditioned Koopman model $(\phi_\theta,K(\theta),\psi_\theta)$.
Comparing its predictions with the true closed loop gives the verification error
$e(\theta)$, which drives the test-and-shrink construction of the region $C$.
The model is then used for one-shot selection of $\widehat\theta(x_0)$ within $C$.}
\label{fig:overview}
\end{figure}

In all of this work the modeled object is either one fixed autonomous system,
or a system driven by a time-varying input that is itself the decision variable
optimized at every step. A third situation is common in engineering practice and
is not covered by either. Many deployed controllers have a fixed structure
together with a small vector of tunable parameters, which must be adapted to the
operating condition. Gain scheduling has been
standard practice in flight and automotive control since long before it received
a systematic treatment~\cite{rugh2000gain}. In modern robotics, locomotion controllers are retuned for different gaits, payloads and terrain, sometimes directly on hardware~\cite{widmer2023tuning}. Safe Bayesian optimization adapts controller parameters online under stability and safety constraints~\cite{sui2015safeopt,dorschel2021safe}, and data-driven tuning has
been extended to high-dimensional embodied
control~\cite{wei2025highdimensional} and to the fine-tuning of learned
policies~\cite{pmlr-v258-tokmak25aPACSBO}.

We denote this parameter vector by $\theta$. Each value of $\theta$ gives
one controller, and closing the loop with it gives one autonomous closed-loop
map $F_\theta$. Once $\theta$ is chosen, it stays fixed for the rest of the
episode. A poor choice degrades tracking, destabilizes the loop,
or, on physical hardware, becomes a safety hazard, and a good one has to be
found quickly. Scoring a candidate $\theta$ on the true nonlinear closed loop
requires a full rollout, which is expensive to repeat and impractical to embed
inside an online search. We therefore seek a differentiable surrogate
that predicts how the closed-loop trajectory changes as $\theta$ varies.

Applying a conventional Koopman model to this problem means training one model
per sampled parameter. The result is a bank of disconnected models, which gives no
prediction at an untested $\theta$, needs a new model for every new controller,
and cannot be optimized continuously over $\theta$. The closest prior work, the
parametric Koopman neural network (PKNN) of~\cite{Guo_parametric_Koopman}, removes the
disconnection by learning one shared dictionary together with a
parameter-indexed family of operators $K(\theta)$. Only the latent
dynamics then vary, while the coordinates stay common.

Shared coordinates matter because exact finite-dimensional Koopman closure is often impossible. This limitation is structural and does not come from
training. When the dynamics admit no finite-dimensional invariant subspace
containing the state, no amount of training produces one, and the residual of
any fixed dictionary is bounded below by an orthogonal projection error. Closed-loop maps with several fixed points or richer attractors need not admit such a subspace at all~\cite{Brunton2016InvariantSubspaces}, and even when an exact closure exists it can be fragile, as the slow-manifold benchmark of Section~\ref{subsubsec:slow_manifold} shows. In the parameterized setting the
difficulty compounds, because the observables that close the representation may
themselves depend on $\theta$. Changing the feedback law changes $F_\theta$,
and with it the local linearization and the transient geometry, so one dictionary is
unlikely to remain informative across the whole family.
Proposition~\ref{prop:projection_floor} makes this precise by exhibiting a
closure-error floor that persists for a fixed dictionary however expressive the
map $K(\cdot)$ is made.
Remark~\ref{rem:strict_separation} gives a pair of closed loops for which a
conditioned dictionary is exact while no shared dictionary of the same size is.

We therefore condition the complete finite-dimensional representation. The
encoder, the latent operator and the decoder all take $\theta$ as an input, so
that both the observable coordinates and the latent linear dynamics vary
continuously with the feedback parameter. The model is trained once, offline, over a
distribution of feedback parameters and initial conditions, and defines a single
surrogate for the entire closed-loop family that is differentiable in
$\theta$ almost everywhere.

The learned representation cannot be expected to be accurate for every admissible feedback parameter. The training data are finite and, more fundamentally, for some controllers the closed loop may have no good finite-dimensional linear description at all. We therefore pair the model with
a test-and-shrink procedure that finds a deployment region $C\subseteq\Theta$
on which the surrogate has passed a prescribed finite-horizon prediction test on
held-out samples. As an example of what the representation enables, we use
it for one-shot controller selection inside $C$. The parameter is optimized once
at the observed initial state, and is then
held fixed for the rest of the episode while the controller keeps acting on state
feedback. This differs from Koopman-based model predictive
control (MPC)~\cite{Korda2018KoopmanMPC,cibulka2025dictionaryfree}, which solves a
new problem after every measurement, and from sequential controller
tuning~\cite{sui2015safeopt,widmer2023tuning}, which needs a new trial for
every candidate.

\smallskip
\noindent\textbf{Contributions.}
(i) We formulate Koopman representation learning for a family of
closed-loop systems parameterized by a feedback control policy (Fig.~\ref{fig:overview}), and distinguish
it from Koopman models whose scheduling variable is a time-varying input.
(ii) We introduce a fully policy-conditioned representation in which the
encoder, Koopman operator, and decoder all depend on the feedback parameter,
with an equilibrium-anchoring construction that fixes the physical equilibrium
at the latent origin for every $\theta$.
(iii) We give a finite-horizon error analysis and a parameter-continuity
result guaranteeing that a minimizer of the surrogate objective exists on a
compact region. We also give an $L^2$ projection bound showing why a shared dictionary
can be less expressive than a conditioned one, with an explicit example where
the difference is strict.
(iv) We develop an empirical verification pipeline that identifies a parameter
region on which held-out samples pass a prescribed finite-horizon prediction test, and
demonstrate multi-step prediction and one-shot feedback-parameter selection on
analytic and MuJoCo benchmarks.\footnote{The code and the trained models are
available at \url{https://github.com/kirangkrishnan/Parameterized-Koopman}.}

\section{RELATED WORK}
\label{sec:related_work}

\subsection{DATA-DRIVEN KOOPMAN APPROXIMATION AND LEARNED REPRESENTATIONS}
The performance of EDMD and Kernel-EDMD~\cite{Williams2015EDMD,williams2015kernel} depends on the choice of dictionary or kernel, since a small one may fail to capture an approximately invariant
subspace, while a very large one raises computational cost and data
requirements. This choice becomes acute when $\theta$ varies, since
one fixed dictionary would have to remain informative for every Koopman operator in a whole family.

Learning the observables removes this manual choice. In~\cite{takeishi2017learning}, for instance, the observables are trained so that a linear map fitted by least squares on the
lifted data leaves a small residual, which makes their span approximately
Koopman invariant. Autoencoders with latent linear dynamics
and multi-step losses recover interpretable embeddings, including for systems
with continuous spectrum~\cite{Lusch2018}. Related formulations learn
low-dimensional linearly recurrent coordinates while managing the trade-off
between expressiveness and overfitting~\cite{Otto2019LRAE,yeung2019learning,Li2017EDMDDL}.

Recursive prediction is the main difficulty, because one-step discrepancies
accumulate when a learned operator is applied repeatedly. Multi-step losses
address this by training on the same recursion used at inference, and a
consistent Koopman autoencoder that exploits forward and backward dynamics
improves forecasting further~\cite{azencot20a_ConsistentKoopman}. Other work
periodically re-encodes predicted states to correct drift~\cite{fathi2024course}.
Our rollouts avoid re-encoding, applying the learned matrix
recursively with no access to intermediate true states, so that the reported
errors reflect the quantity the deployment optimizer evaluates.

These methods all model a single fixed autonomous system. Applied independently at several feedback parameters, they produce a
collection of disconnected models and define no differentiable surrogate at an
unseen parameter.

\subsection{KOOPMAN MODELS WITH CONTROL, AND PARAMETRIC OPERATORS}
DMD with control separates the autonomous dynamics from the effect of measured
inputs~\cite{proctor2016dmdc}; EDMD-based linear predictors have been embedded
in model predictive control~\cite{Korda2018KoopmanMPC}, with later formulations
reducing dictionary dependence while retaining a convex online
problem~\cite{cibulka2025dictionaryfree}. The Koopman Control Family provides
universal finite-dimensional lifted-linear, bilinear, and linear-switched forms
for discrete-time systems with inputs~\cite{haseli2026kcf}, and switched
Koopman models over discrete input sets require a search whose worst-case complexity grows exponentially with the
horizon~\cite{BlischkeHespanha2023}. These predictors retain the input
as a time-varying exogenous variable and are suited to receding-horizon
optimization of an input sequence, which is not the present problem.

Parametric model reduction addresses prediction when a system depends on
physical, geometric, or design parameters, typically by interpolating modal
quantities or reduced operators across parameter
values~\cite{andreuzzi2023parametric,huhn2023parametric}. Interpolating
separately identified models does not, however, learn a common nonlinear
coordinate system jointly with the parameter-to-dynamics map. The closest prior
architecture is PKNN~\cite{Guo_parametric_Koopman} (see also
\cite{PVKO_Parametric_Koopman}), which jointly learns a common dictionary and a
parameter-indexed operator family. It therefore defines a continuous
surrogate, but its coordinates are shared across the family, a restriction that our architecture removes. The conditioning variable also has a different meaning here. Parametric
reduced-order models usually treat the parameter as a plant or operating-condition
variable, whereas $\theta$ parameterizes the feedback law itself. Varying
it changes $F_\theta$, its linearization, and the part of the state space on which
the closed loop respects the operating constraints.

\subsection{DATA-DRIVEN TUNING AND FINITE-SAMPLE VERIFICATION}
Tuning the parameters of a feedback controller can be posed as black-box optimization when closed-loop
performance is measurable but an analytic model is not available. Safe Bayesian
optimization builds probabilistic models of performance and constraints while
sequentially selecting parameters~\cite{sui2015safeopt}, with control-oriented
extensions for robust stable regions~\cite{dorschel2021safe}, hardware
locomotion tuning~\cite{widmer2023tuning}, policy
fine-tuning~\cite{pmlr-v258-tokmak25aPACSBO}, and high-dimensional embodied
parameters~\cite{wei2025highdimensional}. These methods improve the controller
through a sequence of new trials. In our setting, closed-loop data are collected offline to train and test a parameter-conditioned surrogate.
A constrained optimization over the verified parameter region is then solved
just once, and the optimal parameters are held fixed from then on. The verification step is
related to scenario-based optimization, where sampled realizations yield
probabilistic statements about constraint violation~\cite{campi2018scenario}.
Such guarantees require a sampling argument matched to the construction of the
decision set. Our test-and-shrink procedure only reports the region on which
held-out prediction and admissibility tests were passed.

\section{PROBLEM FORMULATION}
\label{sec:preliminaries}

\subsection{NOTATION}
\label{subsec:notation}
For a vector $v$, $\|v\|_2$ is the Euclidean norm and
$\|v\|_Q^2:=v^\top Qv$ for $Q\succeq0$. For a matrix $A$, $\|A\|_F$ and
$\|A\|_2$ are the Frobenius and induced spectral norms, $A^\dagger$ the
Moore--Penrose pseudoinverse, $\sigma_{\min}(A)$ the smallest singular value,
and $F^k$ the $k$-fold composition of a map $F$ with $F^0=\mathrm{id}$. We write
\begin{equation*}
  x_k\in\mathcal X\subseteq\mathbb R^{n},\quad
  u_k\in\mathcal U\subseteq\mathbb R^{m},\quad
  \theta\in\Theta\subseteq\mathbb R^{p},\quad
  z_k\in\mathbb R^{d_z}
\end{equation*}
for the physical state, control input, feedback parameter, and learned latent
state. Unless stated otherwise $\mathcal X$ and $\Theta$ are compact sets
relevant to training, verification, and deployment; no claim is made outside
them. Regulation is to a common equilibrium; when it is nonzero the shift
$\widetilde x=x-x_{\mathrm{ref}}$ is applied, so the paper is written in
equilibrium-centered coordinates with $x_{\mathrm{ref}}=0$.

\subsection{PARAMETERIZED CLOSED-LOOP SYSTEMS}
\label{subsec:param_closed_loop_systems}
Consider the discrete-time plant $x_{k+1}=f(x_k,u_k)$ with
$f:\mathcal X\times\mathcal U\to\mathcal X$; a continuous-time plant is included
through a fixed sampling period and flow discretization. The plant is controlled
by a finite-dimensional family of state-feedback policies
$u_k=\pi_\theta(x_k)$, $\theta\in\Theta$. We call $\Theta$ the admissible
parameter set. In every benchmark it is a product of intervals, and the
deployment region $C$ of Section~\ref{sec:surrogate_application} is a subset of
it. A representative family, used in this
work, is linear in the parameter vector,
\begin{equation}
  \pi_\theta(x)=-\sum_{i=1}^p[\theta]_i\,g_i(x),
  \label{eq:linear_in_parameter_policy}
\end{equation}
with prescribed basis functions $g_i:\mathcal X\to\mathbb R^m$ that may be
linear, polynomial, or trigonometric. The framework is not restricted to
policies linear in $\theta$, and $\pi_\theta$ may be any finite-dimensional
parameterized feedback map for which closed-loop trajectories can be generated,
including a neural policy. We use low-dimensional, physically interpretable
parameterizations because they make the feasible region, equilibrium condition,
and subsequent optimization explicit, isolating the representation-learning
problem from that of optimizing a high-dimensional policy network.

Fixing $\theta$ and closing the loop defines the autonomous map
\begin{equation}
  F_\theta(x):=f\!\left(x,\pi_\theta(x)\right),
  \qquad x_{k+1}=F_\theta(x_k),
  \label{eq:closed_loop_map}
\end{equation}
which is held fixed for the duration of a rollout. Changing $\theta$ therefore
changes the autonomous closed-loop map itself, and the object to be represented
is the family $\mathcal F:=\{F_\theta:\theta\in\Theta\}$. This is different from Koopman models with inputs, which
keep $u_k$ as a time-varying exogenous signal and learn
$z_{k+1}\approx Az_k+Bu_k$ or a bilinear
variant~\cite{Proctor2018KoopmanControl,Korda2018KoopmanMPC}. Here the input is
generated by the policy, and we learn a family of autonomous closed-loop
models indexed by $\theta$. For an initial state
$x_0$ we write $x_k^\theta(x_0):=F_\theta^k(x_0)$ and
$u_k^\theta(x_0):=\pi_\theta(x_k^\theta(x_0))$.

We take the regulation target to be the origin and assume the unforced plant has
an equilibrium there, $f(0,0)=0$, and use an equilibrium-preserving policy family
with $\pi_\theta(0)=0$ for every $\theta$. A raw policy
$\widetilde\pi_\theta$ that violates this can be centered as
$\pi_\theta(x)=\widetilde\pi_\theta(x)-\widetilde\pi_\theta(0)$. It then follows
that $F_\theta(0)=f(0,\pi_\theta(0))=f(0,0)=0$ for all $\theta\in\Theta$, which
establishes a common physical equilibrium for the whole family. A common equilibrium does not
imply that every member is stable.

\subsection{KOOPMAN OPERATORS AND POLICY-CONDITIONED COORDINATES}
\label{subsec:koopman_preliminaries}
Let $\mu$ be a fixed finite reference measure on $\mathcal X$ and let the
ambient observable space be the real Hilbert space
$\mathcal G:=L^2(\mathcal X,\mu)$, a standard choice in Koopman formulations for
parametric systems~\cite{Guo_parametric_Koopman}. For each $\theta\in\Theta$,
assuming composition with $F_\theta$ is well defined on $\mathcal G$, the
Koopman operator $\mathcal U_\theta:\mathcal G\to\mathcal G$ is
\begin{equation}
  (\mathcal U_\theta g)(x):=g\!\left(F_\theta(x)\right),
  \label{eq:koopman_operator}
\end{equation}
which is linear in $g$ even though $F_\theta$ is nonlinear in $x$, and satisfies
$(\mathcal U_\theta^k g)(x)=g(F_\theta^k(x))$. It is generally infinite
dimensional, so practical methods seek finite-dimensional subspaces on which its
action is represented by a matrix~\cite{Williams2015EDMD,Brunton2016InvariantSubspaces}.

Let $\Phi=[\varphi_1\ \cdots\ \varphi_{d_z}]^\top$ be a dictionary with span
$\mathcal V_\Phi$. For a fixed $\theta$, $\mathcal V_\Phi$ is Koopman invariant
if $\mathcal U_\theta\mathcal V_\Phi\subseteq\mathcal V_\Phi$, in which case
there is a matrix $K_\theta$ with $\Phi(F_\theta(x))=K_\theta\Phi(x)$ on
$\mathcal X$, and $z_k=\Phi(x_k)$ evolves as $z_{k+1}=K_\theta z_k$. For a
\emph{family} $\{F_\theta\}$, the far stronger requirement is \emph{common
invariance}, $\mathcal U_\theta\mathcal V_\Phi\subseteq\mathcal V_\Phi$ for
\emph{all} $\theta\in\Theta$; invariance for one map and common invariance over
the family are distinct conditions. Common-invariance concepts and related
finite-dimensional control-family representations are studied
in~\cite{haseli2026kcf}, and parametric models built on a parameter-independent
span in~\cite{Guo_parametric_Koopman}. Exact invariant subspaces that also
permit state reconstruction are restrictive for general nonlinear
systems~\cite{Brunton2016InvariantSubspaces,Brunton2022ModernKoopman}. Given
snapshot pairs $y_j=F_\theta(x_j)$, $j=1,\dots,M$, EDMD instead computes the
least-squares matrix
$K_{\mathrm{EDMD}}\in\arg\min_A\sum_j\|\Phi(y_j)-A\Phi(x_j)\|_2^2
= Y_\Phi X_\Phi^\dagger$, where $X_\Phi$ and $Y_\Phi$ collect the lifted
snapshots columnwise~\cite{Williams2015EDMD}.

In the proposed representation the observable coordinates themselves depend on the feedback parameter, which removes the need for one commonly invariant dictionary. For each $\theta$, the components of the
encoder $\phi_\theta:\mathcal X\to\mathbb R^{d_z}$ span a finite-dimensional
subspace $\mathcal V_\theta\subset\mathcal G$, and this subspace is allowed to
change with $\theta$. The ambient space $\mathcal G$ stays the same, and only the coordinates used to describe each closed loop change. An exact
policy-conditioned representation would satisfy
\begin{equation}
  \phi_\theta\!\left(F_\theta(x)\right)=K(\theta)\phi_\theta(x),
  \qquad
  \psi_\theta\!\left(\phi_\theta(x)\right)=x
  \label{eq:exact_conditioned}
\end{equation}
for all $(x,\theta)\in\mathcal X\times\Theta$, with
$\phi_\theta:\mathcal X\to\mathbb R^{d_z}$,
$K(\theta)\in\mathbb R^{d_z\times d_z}$, and
$\psi_\theta:\mathbb R^{d_z}\to\mathcal X$. For each fixed $\theta$ the latent
evolution is linear; across the family both the coordinates and the matrix vary.
In a learned model the equalities in \eqref{eq:exact_conditioned} hold only
approximately, so we define the one-step equivariance residual
$r_\theta(x):=\phi_\theta(F_\theta(x))-K(\theta)\phi_\theta(x)$ and the
reconstruction residual $d_\theta(x):=\psi_\theta(\phi_\theta(x))-x$. The learned
$k$-step latent and decoded predictions from $x_0$ are
\begin{equation}
  z_k^\theta:=K(\theta)^k\phi_\theta(x_0),
  \qquad
  \widehat x_k^\theta:=\psi_\theta\!\left(z_k^\theta\right),
  \label{eq:conditioned_rollout}
\end{equation}
with physical error $e_k(\theta,x_0):=\widehat x_k^\theta-F_\theta^k(x_0)$.

\subsection{ONE-SHOT FEEDBACK-PARAMETER SELECTION}
\label{subsec:one_shot_problem}
Let $H_{\mathrm{sel}}$ be the finite selection horizon. For an initial state
$x_0$, the true closed-loop cost is
\begin{equation}
  J_{\mathrm{true}}(\theta;x_0):=
  \sum_{k=0}^{H_{\mathrm{sel}}-1}
  L\!\left(x_{k+1}^\theta(x_0),\,u_k^\theta(x_0)\right),
  \label{eq:true_cost}
\end{equation}
with the stage cost $L(x,u)=\|x\|_{Q_x}^2+\|u\|_{R_u}^2$,
$Q_x,R_u\succeq0$, where $x_k^\theta(x_0)=F_\theta^k(x_0)$ is the closed-loop
state and $u_k^\theta(x_0)=\pi_\theta(x_k^\theta(x_0))$ the corresponding input,
as defined in Section~\ref{subsec:param_closed_loop_systems}. Ideally,
the parameter would be chosen by minimizing this cost over a region
$C\subseteq\Theta$,
$\theta^{\mathrm{opt}}(x_0)\in\arg\min_{\theta\in C}J_{\mathrm{true}}(\theta;x_0)$,
which has a minimizer when $C$ is compact and $J_{\mathrm{true}}$ is
continuous. Here $C$ is the deployment region built in
Section~\ref{sec:surrogate_application}, that is, a region on whose held-out samples
the learned model has passed a finite-horizon prediction test. Being in
$C$ says that the model predicts well there. It does not, on its own, say that
the corresponding controller stabilizes the plant.

The difficulty is that $J_{\mathrm{true}}$ depends on the closed-loop trajectory
$x_k^\theta(x_0)$, which in general has no closed form. To evaluate the cost at
one candidate $\theta$, the plant (or the simulator) has to be run under
$\pi_\theta$ for $H_{\mathrm{sel}}$ steps, and an optimizer would repeat this
for every candidate it visits. The learned representation avoids this by
supplying a surrogate cost $\widehat J$ built from the predicted trajectory
\eqref{eq:conditioned_rollout}, giving
\begin{equation}
  \widehat\theta(x_0)\in\arg\min_{\theta\in C}\widehat J(\theta;x_0).
  \label{eq:surrogate_selection}
\end{equation}
Both latent-trajectory and decoded physical-state instantiations of
$\widehat J$ are studied. In either case the optimization variable is the
feedback-parameter vector $\theta$, not a sequence of future inputs. After the
optimization is performed once at the initial condition, the returned parameter
is held fixed, $u_k=\pi_{\widehat\theta(x_0)}(x_k)$ for $k=0,1,\dots$. Thus
``one-shot'' refers to a single feedback-parameter selection event at the start
of the episode; the optimizer itself may evaluate the objective many times.

The goal is therefore one learned model
$\mathcal M=(\phi_\theta,K(\theta),\psi_\theta)$ that meets three
requirements. It should predict finite-horizon closed-loop trajectories at unseen
feedback parameters and initial conditions. It should let us identify, from
data, a deployment region $C$, as large as possible, on which these predictions
meet the prescribed accuracy and admissibility criteria. Finally, it should
support the low-dimensional surrogate optimization
\eqref{eq:surrogate_selection}, after which the selected feedback is deployed
unchanged.

\section{POLICY-CONDITIONED REPRESENTATION LEARNING}
\label{sec:methodology}

\begin{figure*}[t]
\centering
\includegraphics[width=\linewidth]{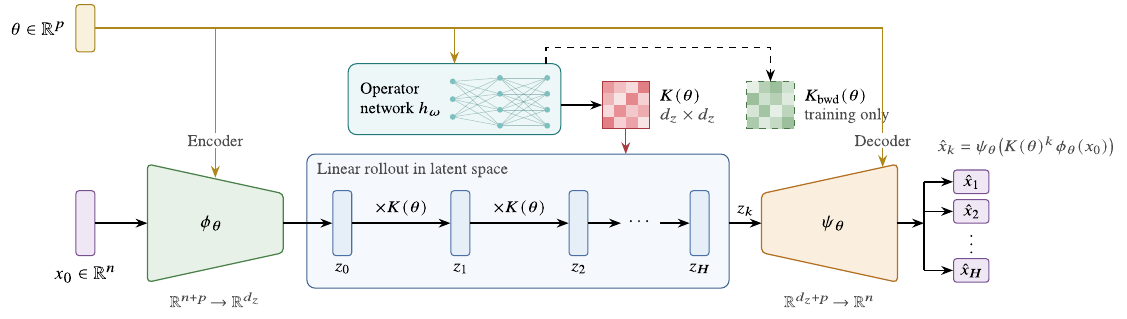}
\caption{Policy-conditioned Koopman representation of a family of closed-loop
systems. The feedback parameter $\theta$ enters the
encoder $\phi_\theta$, the operator network $h_\omega$ that generates
$K(\theta)$, and the decoder $\psi_\theta$. For each fixed $\theta$, prediction
proceeds by repeated application of the forward matrix $K(\theta)$ to the
encoded initial state. The auxiliary backward matrix is used only during
training.}
\label{fig:architecture}
\end{figure*}

\subsection{FROM PER-PARAMETER MODELS TO FULL CONDITIONING}
\label{subsec:conditioned_representation}
A finite-dimensional Koopman model approximates nonlinear state evolution by
linear propagation in a learned observable space. An encoder maps the state to a
latent vector, a matrix propagates it, and a decoder reconstructs the
state~\cite{Lusch2018}. Conventional Koopman modeling addresses a single
dynamical system, which in our setting means a single fixed controller. Since
the goal is to model the family $\{F_\theta\}_{\theta\in\Theta}$, the
representation must vary with the controller parameter.

The direct approach trains a separate model at each sampled parameter
$\theta^{(i)}$:
\begin{equation}
  z_k^{(i)}=\phi_i(x_k),\quad
  z_{k+1}^{(i)}=K_iz_k^{(i)},\quad
  \hat x_k=\psi_i(z_k^{(i)}),
  \label{eq:unconditioned}
\end{equation}
where $\phi_i$, $K_i$, and $\psi_i$ are that model's encoder, Koopman matrix,
and decoder. This approach is accurate at the sampled parameters but produces
a collection of disconnected models. Such a collection defines no continuous surrogate between
training parameters, requires a new model whenever an additional parameter is
considered, and is computationally wasteful.

A scalable alternative retains a parameter-independent dictionary while
conditioning only the latent operator,
\begin{equation}
  z_k=\phi(x_k),\quad
  z_{k+1}=K(\theta)z_k,\quad
  \hat x_k=\psi(z_k),
  \label{eq:operator_conditioned}
\end{equation}
so that the latent dynamics vary continuously with
$\theta$~\cite{Guo_parametric_Koopman}. The encoder and decoder remain fixed, however, and thus every closed-loop system in the family must share one set of
observable coordinates. We instead condition the complete representation,
\begin{equation}
  z_k^\theta=\phi_\theta(x_k),\quad
  z_{k+1}^\theta=K(\theta)z_k^\theta,\quad
  \hat x_k^\theta=\psi_\theta(z_k^\theta),
  \label{eq:fully_conditioned}
\end{equation}
giving the $k$-step prediction
$\hat x_k^\theta=\psi_\theta(K(\theta)^k\phi_\theta(x_0))$. In our implementation the coordinate maps take the parameter as part of their
input, so that $\phi_\theta:\mathbb R^{n+p}\to\mathbb R^{d_z}$ acts on
$[x;\theta]$ and $\psi_\theta:\mathbb R^{d_z+p}\to\mathbb R^{n}$ acts on
$[z;\theta]$, as in Fig.~\ref{fig:architecture}.
Both the latent dynamics and the coordinates in which they are expressed may
change with the controller, and the surrogate spans a continuous
family of closed-loop systems. Section~\ref{subsec:results_ablation} isolates the
contribution of each conditioning decision.

\subsection{ARCHITECTURE}
\label{subsec:architecture}
As shown in Fig.~\ref{fig:architecture}, the model has three jointly trained
components. The \emph{encoder} and \emph{decoder} are multilayer perceptrons
with $\tanh$ activations, each anchored at the equilibrium by subtracting
its own value at the origin:
\begin{align}
  \phi_\theta(x)&=\mathrm{MLP}_{\mathrm{enc}}([x;\theta])-\mathrm{MLP}_{\mathrm{enc}}([0;\theta]),
  \label{eq:encoder}\\
  \psi_\theta(z)&=\mathrm{MLP}_{\mathrm{dec}}([z;\theta])-\mathrm{MLP}_{\mathrm{dec}}([0;\theta]).
  \label{eq:decoder}
\end{align}
This enforces $\phi_\theta(0)=0$ and $\psi_\theta(0)=0$ for every
$\theta\in\Theta$, so the physical equilibrium is represented by the latent
origin for every feedback parameter. The anchoring removes parameter-dependent latent
offsets at the equilibrium and reduces artificial drift during repeated latent
propagation. Since the decoder is continuous, it also ensures that convergence of
the predicted latent state to zero implies convergence of the decoded state to
the physical equilibrium.

The \emph{operator network} $h_\omega:\mathbb R^p\to\mathbb R^{2d_z^2}$, a
multilayer perceptron with ReLU activations, maps the
feedback parameter to forward and backward matrices. Its output is partitioned
as $h_\omega(\theta)=[h_{\mathrm{fwd}}(\theta);h_{\mathrm{bwd}}(\theta)]$ with
$h_{\mathrm{fwd}},h_{\mathrm{bwd}}\in\mathbb R^{d_z^2}$, and each block is
reshaped into a $d_z\times d_z$ matrix $K_{\mathrm{fwd}}(\theta)$,
$K_{\mathrm{bwd}}(\theta)$. The backward matrix is used only to build auxiliary
backward-prediction and consistency losses during training; at prediction and
deployment only the forward matrix is required, and we write
$K(\theta):=K_{\mathrm{fwd}}(\theta)$ throughout.

Given $x_0$ and $\theta$, the latent trajectory is
$z_0^\theta=\phi_\theta(x_0)$ and
\begin{equation}
  z_k^\theta=K(\theta)^kz_0^\theta,
  \qquad
  \hat x_k^\theta=\psi_\theta(z_k^\theta),
  \qquad k=1,\dots,H.
  \label{eq:latent_rollout}
\end{equation}
The same matrix is applied at every step because the feedback parameter is fixed
during the predicted rollout. Nonlinear closed-loop evolution in state space is thus replaced by a linear rollout in latent space, and nonlinearity enters only through the coordinate maps and the dependence of $K$ on $\theta$.

\subsection{CLOSED-LOOP DATA GENERATION}
\label{subsec:data_generation}
Training uses finite-horizon closed-loop trajectories, and the data therefore sample the autonomous map \eqref{eq:closed_loop_map} directly. For
the $i$th trajectory we sample $\theta^{(i)}\sim\mathrm{Uniform}(\Theta)$ and
$x_0^{(i)}\sim\mathrm{Uniform}(\mathcal X_{\mathrm{IC}})$, where
$\mathcal X_{\mathrm{IC}}$ is the training initial-condition set, and roll out
$H$ steps to obtain
$\tau^{(i)}=(x_0^{(i)},\theta^{(i)},x_1^{(i)},\dots,x_H^{(i)})$, giving the dataset
$\mathcal D=\{\tau^{(i)}\}_{i=1}^N$. Training and validation sets use the same
rollout horizon $H$ as the prediction horizon in the loss.

\subsection{TRAINING OBJECTIVE}
\label{subsec:training_objective}
The objective mirrors how the surrogate is used at deployment, where the state is encoded, propagated
repeatedly with $K(\theta)$ and decoded. It combines four terms that enforce accurate
reconstruction, accurate decoded multi-step prediction, approximately linear
latent evolution, and forward--backward consistency, and a fifth that penalizes the operator norm.

The \emph{reconstruction} loss
$\mathcal L_{\mathrm{rec}}=\|\psi_\theta(\phi_\theta(x_0))-x_0\|_2^2$
prevents the encoder from mapping distinct states to indistinguishable latent
vectors. The \emph{prediction} loss is the primary state-space accuracy term, since the
deployment optimizer scores candidates using this rollout:
\begin{align}
  \mathcal L_{\mathrm{pred}}^{\mathrm{fwd}}
  &=\frac1H\sum_{k=1}^{H}
  \left\|\psi_\theta\!\left(K_{\mathrm{fwd}}(\theta)^k\phi_\theta(x_0)\right)-x_k\right\|_2^2,
  \label{eq:loss_pred_fwd}\\
  \mathcal L_{\mathrm{pred}}^{\mathrm{bwd}}
  &=\frac1H\sum_{k=0}^{H-1}
  \left\|\psi_\theta\!\left(K_{\mathrm{bwd}}(\theta)^{H-k}\phi_\theta(x_H)\right)-x_k\right\|_2^2,
  \label{eq:loss_pred_bwd}
\end{align}
with $\mathcal L_{\mathrm{pred}}=\mathcal L_{\mathrm{pred}}^{\mathrm{fwd}}
+\mathcal L_{\mathrm{pred}}^{\mathrm{bwd}}$. The backward term measures whether
the same representation recovers earlier states from the final one. The benefit
of exploiting backward dynamics in Koopman forecasting has been shown
in~\cite{azencot20a_ConsistentKoopman}, and we adapt it by learning both
operators as functions of $\theta$.

Because decoded accuracy alone does not ensure that encoded trajectories are closed under a linear latent evolution, the \emph{latent-linearity} loss
$\mathcal L_{\mathrm{lin}}=\mathcal L_{\mathrm{lin}}^{\mathrm{fwd}}
+\mathcal L_{\mathrm{lin}}^{\mathrm{bwd}}$ compares propagated latent states with
encodings of the corresponding true states:
\begin{align}
  \mathcal L_{\mathrm{lin}}^{\mathrm{fwd}}
  &=\frac1H\sum_{k=1}^{H}\left\|K_{\mathrm{fwd}}(\theta)^k\phi_\theta(x_0)-\phi_\theta(x_k)\right\|_2^2,
  \label{eq:loss_lin_fwd}\\
  \mathcal L_{\mathrm{lin}}^{\mathrm{bwd}}
  &=\frac1H\sum_{k=0}^{H-1}\left\|K_{\mathrm{bwd}}(\theta)^{H-k}\phi_\theta(x_H)-\phi_\theta(x_k)\right\|_2^2.
  \label{eq:loss_lin_bwd}
\end{align}
The backward prediction loss trains $K_{\mathrm{bwd}}(\theta)$ but does not tie
it to $K_{\mathrm{fwd}}(\theta)$, so following~\cite{azencot20a_ConsistentKoopman}
we penalize departure from an inverse relationship with
$\mathcal L_{\mathrm{cons}}=\|K_{\mathrm{fwd}}(\theta)K_{\mathrm{bwd}}(\theta)-I_{d_z}\|_F^2$.
This term acts as a soft regularizer rather than a hard reversibility constraint, because
some closed-loop systems are dissipative or only approximately reversible over
the finite horizon. Finally,
$\mathcal L_{\mathrm{reg}}=\|K_{\mathrm{fwd}}(\theta)\|_F^2+\|K_{\mathrm{bwd}}(\theta)\|_F^2$
discourages large operator norms, which would amplify latent errors under the
powers $K_{\mathrm{fwd}}(\theta)^k$; it improves conditioning but does not by
itself guarantee spectral stability. The total objective is
\begin{equation}
\begin{aligned}
  \mathcal L_{\mathrm{total}}
  &=\lambda_{\mathrm{rec}}\mathcal L_{\mathrm{rec}}
  +\lambda_{\mathrm{pred}}\mathcal L_{\mathrm{pred}}
  +\lambda_{\mathrm{lin}}\mathcal L_{\mathrm{lin}}\\
  &\quad+\lambda_{\mathrm{cons}}\mathcal L_{\mathrm{cons}}
  +\lambda_{\mathrm{reg}}\mathcal L_{\mathrm{reg}},
\end{aligned}
  \label{eq:total_loss}
\end{equation}
minimized over the trainable parameters $\eta$ of the encoder, decoder, and
operator network as
$\min_\eta|\mathcal B|^{-1}\sum_{\tau\in\mathcal B}\mathcal L_{\mathrm{total}}(\tau;\eta)$
for minibatches $\mathcal B\subseteq\mathcal D$. The weights
$\lambda_{(\cdot)}$ balance the five terms; their values are given in Section~\ref{subsec:benchmarks}.

\section[FINDING THE DEPLOYMENT REGION AND SELECTING THE CONTROLLER]{FINDING THE DEPLOYMENT REGION AND SELECTING THE CONTROLLER}
\label{sec:surrogate_application}

The learned model gives an almost-everywhere differentiable surrogate for the closed-loop behavior
at each feedback parameter, but it cannot be trusted everywhere in
$\Theta$. Its accuracy varies across the parameter set because it is fitted to finitely many trajectories. In addition, for some controllers a linear description in a $d_z$-dimensional latent space may not exist, and more data or longer training cannot change this (Section~\ref{subsec:theory_projection}). We therefore only select controllers
from a region on which the final surrogate has passed a finite-horizon
prediction test, and call it the \emph{deployment region}. The deployment
region is part of the method itself, since any use of the surrogate should be confined to it. The one-shot controller selection described below is one example of
such a use, included to show how the representation can be applied. The trained model is
used through the offline-to-online sequence
$\mathcal M_{\mathrm{scout}}\to C_{\mathrm{pre}}\to\mathcal M_{\mathrm{final}}
\to C\to\theta_s^\star$, whose offline part is summarized in Algorithm~\ref{alg:offline}.

\subsection{EMPIRICALLY VERIFIED REGION CONSTRUCTION}
\label{subsec:verification}
To test a candidate parameter, we fix a finite set of initial conditions
$\mathcal X_{\mathrm{ver}}$ inside the training initial-condition set $\mathcal X_{\mathrm{IC}}$. We simulate the true closed loop from each of
them and compare the result with the prediction of the model. The verification
error of $\theta$ is the worst relative error over these initial conditions and
over the horizon,
\begin{equation}
  e(\theta)=
  \max_{\substack{x_0\in\mathcal X_{\mathrm{ver}}\\1\leq k\leq H}}
  \frac{\left\|\psi_\theta\!\left(K(\theta)^k\phi_\theta(x_0)\right)-F_\theta^k(x_0)\right\|_2}
       {\max\!\left(\|F_\theta^k(x_0)\|_2,\delta\right)}\times100,
  \label{eq:cert_error}
\end{equation}
where $\delta>0$ prevents numerical amplification near the equilibrium
($\delta=0.1$ in all experiments). A parameter qualifies if
$e(\theta)\leq\varepsilon_{\max}$ and the corresponding true trajectory remains
inside the operating constraints. Verification proceeds in three phases.

\emph{Phase 1 (exploration).} Sample
$\theta^{(i)}\sim\mathrm{Uniform}(\Theta)$, $i=1,\dots,N_{\mathrm{exp}}$,
simulate the true closed loop from $\mathcal X_{\mathrm{ver}}$, evaluate
\eqref{eq:cert_error}, retain
$G_1=\{\theta^{(i)}:e(\theta^{(i)})\leq\varepsilon_{\max}\}$, and form
$H_1=\operatorname{ConvHull}(G_1)$ as a first approximation of the reliable
region.

\begin{figure}[t]
    \centering
    \includegraphics[width=\columnwidth]{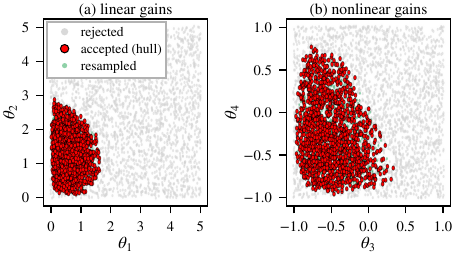}
    \caption{Deployment region for the VdP benchmark. The
four-dimensional parameter samples are shown in two projections: (a)
the linear gains $(\theta_1,\theta_2)$ and (b) the nonlinear gains
$(\theta_3,\theta_4)$. Both panels span the full admissible set
$\Theta_{\mathrm{VdP}}$. Gray points are sampled parameters that were rejected
($e(\theta)>\varepsilon_{\max}$), red points are the accepted parameters that
define the converged hull, and green points are the samples drawn from the
final hull for retraining and verification. The other benchmarks are handled
in the same way.}
    \label{fig:hull_shrinkage}
\end{figure}

\emph{Phase 2 (iterative hull shrinkage).} The hull may still contain untested
interior points at which the surrogate is unreliable. At iteration $t$ we sample
$\theta^{(j)}\sim\mathrm{Uniform}(H_t)$, apply the same error and admissibility
tests to obtain $G_{t+1}$, and set $H_{t+1}=\operatorname{ConvHull}(G_{t+1})$.
This shrinks the region toward the subset where the criterion is consistently
met, as illustrated for the Van der Pol (VdP) benchmark in
Fig.~\ref{fig:hull_shrinkage}. The iteration stops when the hull volume stabilizes,
$\operatorname{vol}(H_{t+1})/\operatorname{vol}(H_t)\geq0.99$, or when the
survival rate, the fraction of samples that pass the test, exceeds the desired threshold.

\emph{Phase 3 (final verification).} The surrogate is retrained on trajectories
concentrated in the hull $C_{\mathrm{pre}}$ from Phase~2, so that accuracy improves precisely in the
part of parameter space the deployment optimizer will search. Hence a scout model is used in Phases~1 and~2 and a final model in Phase~3. We then draw
$\theta^{(j)}\sim\mathrm{Uniform}(C_{\mathrm{pre}})$, $j=1,\dots,N_s$ ($N_s=1500$ in the
experiments), and accept the region if the empirical survival rate is at least $0.95$. Otherwise the
passing samples span a smaller hull, a new batch is drawn from it, and the test
is repeated until the survival rate reaches $0.95$ or too few samples pass to form a hull. The accepted hull is the one from which the last batch was drawn,
so we check its survival rate again on an independent draw of $1500$
parameters. Uniform sampling over a hull is implemented by
rejection sampling from its axis-aligned bounding box with a Delaunay
point-in-hull test~\cite{Delauney_convex_1,Bowyer1981ComputingDT_Delauney};
since the proposal density is constant, conditioning on acceptance is uniform
over the hull. The accepted region, which we also call the verified region, has passed the stated finite-sample
procedure. This does not imply that every untested parameter in the hull
satisfies the criterion, absent an additional probabilistic or deterministic
guarantee.

\begin{algorithm}[t]
\caption{Offline training and verified-region construction}
\label{alg:offline}
\begin{algorithmic}[1]
\REQUIRE Admissible parameter set $\Theta$; sets $\mathcal X_{\mathrm{IC}}$, $\mathcal X_{\mathrm{ver}}$; horizon $H$; tolerance $\varepsilon_{\max}$
\REQUIRE Budgets $N_{\mathrm{scout}},N_{\mathrm{final}},N_{\mathrm{exp}},N_{\mathrm{hull}},N_s$
\STATE \textbf{function} \textsc{Train}$(\Omega,N)$:
\STATE \quad \textbf{for} $i=1,\dots,N$ \textbf{do} sample $\theta^{(i)}\!\sim\!\mathrm{Uniform}(\Omega)$, $x_0^{(i)}\!\sim\!\mathrm{Uniform}(\mathcal X_{\mathrm{IC}})$;
\STATE \quad\quad roll out $x_{k+1}^{(i)}=f(x_k^{(i)},\pi_{\theta^{(i)}}(x_k^{(i)}))$ for $k<H$; store $\tau^{(i)}$
\STATE \quad fit $(\phi_\theta,K(\theta),\psi_\theta)$ by minimizing \eqref{eq:total_loss}; \textbf{return} $\mathcal M$
\STATE $\mathcal M_{\mathrm{scout}}\leftarrow\textsc{Train}(\Theta,N_{\mathrm{scout}})$
\STATE \textbf{Phase 1:} sample $\{\theta^{(i)}\}_{i=1}^{N_{\mathrm{exp}}}$ from $\Theta$; evaluate $e(\theta^{(i)})$ by \eqref{eq:cert_error}
\STATE $G_1\leftarrow\{\theta^{(i)}:e(\theta^{(i)})\leq\varepsilon_{\max}\}$;\ \ $H_1\leftarrow\operatorname{ConvHull}(G_1)$;\ \ $t\leftarrow1$
\STATE \textbf{Phase 2:}
\WHILE{hull volume not converged}
  \STATE sample $\{\theta^{(j)}\}_{j=1}^{N_{\mathrm{hull}}}$ from $H_t$; evaluate $e(\theta^{(j)})$
  \STATE $G_{t+1}\leftarrow\{\theta^{(j)}\in H_t:e(\theta^{(j)})\leq\varepsilon_{\max}\}$
  \STATE $H_{t+1}\leftarrow\operatorname{ConvHull}(G_{t+1})$;\ \ $t\leftarrow t+1$
\ENDWHILE
\STATE $C_{\mathrm{pre}}\leftarrow H_t$;\ \ $\mathcal M_{\mathrm{final}}\leftarrow\textsc{Train}(C_{\mathrm{pre}},N_{\mathrm{final}})$
\STATE \textbf{Phase 3:} sample $\{\theta^{(j)}\}_{j=1}^{N_s}$ from $C_{\mathrm{pre}}$; evaluate $e(\theta^{(j)})$ with $\mathcal M_{\mathrm{final}}$
\IF{the prescribed fraction satisfies $e(\theta^{(j)})\leq\varepsilon_{\max}$}
  \STATE $C\leftarrow C_{\mathrm{pre}}$
\ELSE
  \STATE shrink or resample $C_{\mathrm{pre}}$ and repeat Phase 3
\ENDIF
\RETURN $\mathcal M_{\mathrm{final}}$ and verified region $C$
\end{algorithmic}
\end{algorithm}

\subsection{ONE-SHOT SYNTHESIS AND DEPLOYMENT}
\label{subsec:one_shot_synthesis}
At deployment the surrogate selects a single feedback parameter for the observed
initial state $x_0$. For a candidate
$\theta\in C$ the initial state is lifted and propagated by
\eqref{eq:latent_rollout} over the selection horizon $H_{\mathrm{sel}}$. The two
objectives below use the same encoder, operator, decoder, verified region, and latent
rollout, and differ only in the coordinates used to score a candidate.

The \emph{latent-trajectory} objective scores a candidate by the
accumulated magnitude of the predicted latent trajectory,
\begin{equation}
  \widehat J_z(\theta;x_0)=\sum_{k=1}^{H_{\mathrm{sel}}}(z_k^\theta)^\top Q_zz_k^\theta
  =q_z\sum_{k=1}^{H_{\mathrm{sel}}}\left\|K(\theta)^k\phi_\theta(x_0)\right\|_2^2,
  \label{eq:latent_objective}
\end{equation}
where the right-hand form is the reported implementation with
$Q_z=q_zI_{d_z}$. The equilibrium anchoring $\phi_\theta(0)=0$,
$\psi_\theta(0)=0$ ensures that the desired physical equilibrium is the latent
origin for every candidate. The objective \eqref{eq:latent_objective} therefore tests whether the
learned coordinates themselves provide a useful regulation surrogate; we do not
assume the latent magnitude is equivalent to a physical cost. The \emph{decoded}
objective scores the same predicted trajectory in physical coordinates,
\begin{equation}
  \widehat J_x(\theta;x_0)=\sum_{k=1}^{H_{\mathrm{sel}}}
  \left\|\hat x_k^\theta\right\|_{Q_x}^2.
  \label{eq:decoded_objective}
\end{equation}
As in \eqref{eq:latent_objective}, the target is the origin; a nonzero
target is handled by the shift in Section~\ref{subsec:notation}. The trajectory
is still propagated through the linear latent dynamics and decoding is used
only for evaluation, so this remains a low-dimensional surrogate optimization
that never simulates the true plant inside the optimizer.

Representing the verified region as $C=\{\theta:A_C\theta\leq b_C\}$, the
deployed parameter for $s\in\{z,x\}$ solves
\begin{equation}
\begin{aligned}
  \min_{\theta\in\mathbb R^p}\quad&\widehat J_s(\theta;x_0)\\
  \text{s.t.}\quad&A_C\theta\leq b_C,\qquad
  \theta_{\min}\leq\theta\leq\theta_{\max}.
\end{aligned}
\label{eq:constrained_selection}
\end{equation}
The solution $\theta_s^\star$ of \eqref{eq:constrained_selection} is the minimizer $\widehat\theta(x_0)$ of \eqref{eq:surrogate_selection} with $\widehat J=\widehat J_s$. In the reported implementation we solve \eqref{eq:constrained_selection} with
sequential least-squares quadratic programming
(SLSQP)~\cite{kraft1988slsqp}, a sequential quadratic programming method that
handles the linear constraints directly. It generates feasible
iterates from local first-order information, with the objective Jacobian
obtained by automatic differentiation through the encoder, the
parameter-dependent Koopman rollout, and the decoder. The initial point is the
mean of the verified samples unless stated otherwise. Because the operator
network uses ReLU activations, $\widehat J_s$ is continuous and piecewise smooth
in $\theta$. It is differentiable except on the measure-zero set where a ReLU
switches, and automatic differentiation returns a one-sided derivative there.
SLSQP assumes a smooth objective, so we use it as a local method and judge the
selected parameters on the true plant in Section~\ref{sec:experiments}.

Each objective evaluation requires one encoder evaluation and $H_{\mathrm{sel}}$ latent
matrix--vector products, plus as many decoder evaluations for $\widehat J_x$, giving a dominant cost of
$O(H_{\mathrm{sel}}d_z^2)$. The decision variable has dimension $p$, typically far
smaller than the control sequence optimized by MPC. Once selected, $\theta_s^\star$ stays fixed, and the per-step online cost is only that of evaluating $\pi_{\theta_s^\star}(x_t)$.

\section{THEORETICAL ANALYSIS}
\label{sec:theory}

This section works over a finite horizon and a compact region, without
assuming an exact finite-dimensional Koopman linearization. We bound how
reconstruction and closure defects propagate and examine what the forward--backward consistency term guarantees. We then show that the rollout depends
continuously on $\theta$ and explain why a fixed dictionary can leave
a closure error that no choice of the map $K(\cdot)$ removes.

\subsection{FINITE-HORIZON ERROR PROPAGATION}
\label{subsec:theory_finite_horizon}
Fix $\theta\in C$ and write $K_\theta:=K(\theta)$. The recursive predictions are
$\widehat z_k^\theta=K_\theta^k\phi_\theta(x_0)$ and
$\widehat x_k^\theta=\psi_\theta(\widehat z_k^\theta)$, while the true state in
learned coordinates is $z_k^\theta=\phi_\theta(x_k^\theta)$.

\begin{assumption}[Operating sets]
\label{ass:domain}
The parameter region $C\subseteq\Theta$ and the initial-condition set
$\mathcal X_0\subseteq\mathcal X$ are compact and nonempty, and the closed-loop
map $(x,\theta)\mapsto F_\theta(x)$ is continuous on $\mathcal X\times\Theta$,
which holds when $f$ and $(x,\theta)\mapsto\pi_\theta(x)$ are continuous, input
saturation included. The finite-horizon
state set
\begin{equation*}
  \mathcal X_H:=\left\{F_\theta^k(x_0)\ :\ \theta\in C,\ x_0\in\mathcal X_0,\ 0\leq k\leq H\right\}
\end{equation*}
is contained in a compact subset of $\mathcal X$, and the latent set
\begin{equation*}
\begin{aligned}
  \mathcal Z_H:={}&\left\{K(\theta)^k\phi_\theta(x_0)\ :\ \theta\in C,\ x_0\in\mathcal X_0,\ 0\leq k\leq H\right\}\\
  &\cup\left\{\phi_\theta(x)\ :\ \theta\in C,\ x\in\mathcal X_H\right\},
\end{aligned}
\end{equation*}
which contains both the predicted latent states and the encodings of the
true states, is bounded. The maps $\phi$, $\psi$ and $K$ are continuous and locally Lipschitz
on $\mathcal X\times\Theta$, $\mathbb R^{d_z}\times\Theta$ and $\Theta$
respectively, which holds for the feedforward networks used here with the
activations of Section~\ref{subsec:architecture}.
\end{assumption}

Assumption~\ref{ass:domain} already supplies the four constants that the error
analysis needs, and none of them has to be assumed separately. Each of the
quantities below is a continuous function of its arguments on a compact set and
therefore attains a finite maximum there; local Lipschitz continuity on a
compact set gives a finite global Lipschitz constant on that set.

\noindent\textbf{Constants.} Under Assumption~\ref{ass:domain}, define
\begin{align}
  \varepsilon_{\mathrm{rec}}
  &:=\max_{\theta\in C,\,x\in\mathcal X_H}\left\|\psi_\theta(\phi_\theta(x))-x\right\|_2,
  \label{eq:rec_defect}\\
  \varepsilon_{\mathrm{lin}}
  &:=\max_{\theta\in C,\,x\in\mathcal X_H}\left\|\phi_\theta(F_\theta(x))-K(\theta)\phi_\theta(x)\right\|_2,
  \label{eq:closure_defect}\\
  \kappa&:=\max_{\theta\in C}\|K(\theta)\|_2,
  \label{eq:op_lip}\\
  L_\psi&:=\sup_{\substack{\theta\in C,\ z\neq z'\\ z,z'\in\overline{\mathcal Z}_H}}
  \frac{\|\psi_\theta(z)-\psi_\theta(z')\|_2}{\|z-z'\|_2},
  \label{eq:lpsi_def}
\end{align}
with $\overline{\mathcal Z}_H$ the closed convex hull of $\mathcal Z_H$. All
four are finite and nonnegative.

The constants $\varepsilon_{\mathrm{rec}}$ and $\varepsilon_{\mathrm{lin}}$
are the uniform regional counterparts of the reconstruction and latent-linearity
losses minimized during training. They exist under Assumption~\ref{ass:domain}, and training is meant to make them small. The analysis does not show that training succeeds, because small empirical averages over a finite sample do
not by themselves bound a maximum over a continuous region. The bounds below therefore describe how these two errors propagate, and their size remains an empirical question.

\begin{theorem}[Finite-horizon prediction error]
\label{thm:prediction}
Under Assumption~\ref{ass:domain}, with the constants \eqref{eq:rec_defect}--\eqref{eq:lpsi_def}, let
$S_k(\kappa):=\sum_{j=0}^{k-1}\kappa^j$, equal to $(\kappa^k-1)/(\kappa-1)$ for
$\kappa\neq1$ and to $k$ for $\kappa=1$. Then for every $\theta\in C$,
$x_0\in\mathcal X_0$, and $k=1,\dots,H$,
\begin{align}
  \|\widehat z_k^\theta-z_k^\theta\|_2&\leq\varepsilon_{\mathrm{lin}}S_k(\kappa),
  \label{eq:latent_error}\\
  \|\widehat x_k^\theta-x_k^\theta\|_2&\leq\varepsilon_{\mathrm{rec}}
  +L_\psi\varepsilon_{\mathrm{lin}}S_k(\kappa).
  \label{eq:state_error}
\end{align}
\end{theorem}

\begin{proof}
By \eqref{eq:closure_defect} the encoded true trajectory satisfies
$z_{k+1}^\theta=K_\theta z_k^\theta+r_k^\theta$ with
$\|r_k^\theta\|_2\leq\varepsilon_{\mathrm{lin}}$. Let
$e_k^z:=\widehat z_k^\theta-z_k^\theta$. Since
$\widehat z_{k+1}^\theta=K_\theta\widehat z_k^\theta$ and $e_0^z=0$, we have
$e_{k+1}^z=K_\theta e_k^z-r_k^\theta$ and therefore
$\|e_{k+1}^z\|_2\leq\kappa\|e_k^z\|_2+\varepsilon_{\mathrm{lin}}$; iterating gives
\eqref{eq:latent_error}. For the decoded state, adding and subtracting
$\psi_\theta(z_k^\theta)$ and using \eqref{eq:rec_defect}--\eqref{eq:lpsi_def},
\begin{equation*}
  \|\widehat x_k^\theta-x_k^\theta\|_2
  \leq L_\psi\|\widehat z_k^\theta-z_k^\theta\|_2
  +\|\psi_\theta(\phi_\theta(x_k^\theta))-x_k^\theta\|_2,
\end{equation*}
and substituting \eqref{eq:latent_error} proves \eqref{eq:state_error}.
\end{proof}

\begin{remark}[Error growth]
\label{cor:growth}
The latent-closure contribution in \eqref{eq:state_error} is uniformly bounded
in $k$ when $\kappa<1$, grows at most linearly when $\kappa=1$, and may grow
geometrically when $\kappa>1$. In particular, for $\kappa<1$,
$\|\widehat x_k^\theta-x_k^\theta\|_2\leq\varepsilon_{\mathrm{rec}}
+L_\psi\varepsilon_{\mathrm{lin}}/(1-\kappa)$.
\end{remark}

This bound motivates recursive multi-step training, although it is not a stability theorem, since neither a Frobenius penalty on $K_\theta$ nor $\rho(K_\theta)<1$ alone
proves stability of the true nonlinear closed loop. The auxiliary backward
operator acts as a representation-learning regularizer, following consistent
Koopman autoencoders~\cite{azencot20a_ConsistentKoopman}, and Proposition~\ref{prop:consistency} states what it guarantees.

\begin{proposition}[Forward--backward consistency]
\label{prop:consistency}
Fix $\theta\in C$ and let $K_f:=K_{\mathrm{fwd}}(\theta)$,
$K_b:=K_{\mathrm{bwd}}(\theta)$. If $\|K_fK_b-I\|_2\leq\eta<1$, then $K_f$ and
$K_b$ are nonsingular and
\begin{equation}
  \sigma_{\min}(K_f)\geq\frac{1-\eta}{\|K_b\|_2},
  \qquad
  \sigma_{\min}(K_b)\geq\frac{1-\eta}{\|K_f\|_2}.
  \label{eq:sigma_bounds}
\end{equation}
\end{proposition}

\begin{proof}
By the Neumann series, $K_fK_b$ is invertible with
$\|(K_fK_b)^{-1}\|_2\leq(1-\eta)^{-1}$, so both square factors are invertible.
The bounds follow from $K_f^{-1}=K_b(K_fK_b)^{-1}$, $K_b^{-1}=(K_fK_b)^{-1}K_f$
and $\sigma_{\min}(A)=1/\|A^{-1}\|_2$.
\end{proof}

\begin{remark}
\label{rem:consistency}
Proposition~\ref{prop:consistency} is conditional and pointwise in $\theta$. It
shows that a sufficiently small consistency defect keeps the forward
operator nonsingular, so the latent rollout cannot collapse onto a
lower-dimensional subspace; it does not imply that the physical dynamics are reversible,
contracting, or stable.
\end{remark}

\subsection{CONTINUITY IN THE FEEDBACK PARAMETER}
\label{subsec:theory_continuity}
Policy conditioning is intended to produce one surrogate that varies continuously with the feedback parameter, in place of a collection of independently fitted models. We now show that the learned rollout has this property, and hence that the selection problem has a solution.

\begin{assumption}[Regularity in the feedback parameter]
\label{ass:parameter_regularity}
There exist constants $L_K$, $L_{\phi,\theta}$, $L_{\psi,\theta}$ and
$B_\phi$ such that for all $\theta,\vartheta\in C$, all $x\in\mathcal X_H$ and
all $z\in\overline{\mathcal Z}_H$,
$\|K(\theta)-K(\vartheta)\|_2\leq L_K\|\theta-\vartheta\|_2$,
$\|\phi_\theta(x)-\phi_\vartheta(x)\|_2\leq L_{\phi,\theta}\|\theta-\vartheta\|_2$,
$\|\psi_\theta(z)-\psi_\vartheta(z)\|_2\leq L_{\psi,\theta}\|\theta-\vartheta\|_2$,
and $\|\phi_\theta(x)\|_2\leq B_\phi$. As with the constants \eqref{eq:rec_defect}--\eqref{eq:lpsi_def}, these
follow from Assumption~\ref{ass:domain} for the networks used here; they are
collected separately because the bounds of
Proposition~\ref{prop:continuity} are stated in terms of them.
\end{assumption}

\begin{proposition}[Continuous dependence on $\theta$]
\label{prop:continuity}
Under Assumption~\ref{ass:domain} and Assumption~\ref{ass:parameter_regularity}, for
$k=1,\dots,H$,
$\|\widehat z_k^\theta-\widehat z_k^\vartheta\|_2\leq L_{z,k}\|\theta-\vartheta\|_2$
and
$\|\widehat x_k^\theta-\widehat x_k^\vartheta\|_2\leq L_{x,k}\|\theta-\vartheta\|_2$,
where, with $\bar\kappa:=\max\{1,\kappa\}$,
\begin{equation}
  L_{z,k}:=\kappa^kL_{\phi,\theta}+k\bar\kappa^{k-1}L_KB_\phi,
  \qquad
  L_{x,k}:=L_\psi L_{z,k}+L_{\psi,\theta}.
  \label{eq:lipschitz_constants}
\end{equation}
Consequently every continuous finite-horizon surrogate objective attains a
minimum on the nonempty compact region $C$.
\end{proposition}

\begin{proof}
The identity $A^k-B^k=\sum_{j=0}^{k-1}A^{k-1-j}(A-B)B^j$ gives
$\|K_\theta^k-K_\vartheta^k\|_2\leq k\bar\kappa^{k-1}L_K\|\theta-\vartheta\|_2$.
Adding and subtracting $K_\theta^k\phi_\vartheta(x_0)$,
\begin{align*}
  \|\widehat z_k^\theta-\widehat z_k^\vartheta\|_2
  &\leq\|K_\theta^k\|_2\|\phi_\theta(x_0)-\phi_\vartheta(x_0)\|_2\\
  &\quad+\|K_\theta^k-K_\vartheta^k\|_2\|\phi_\vartheta(x_0)\|_2,
\end{align*}
which yields the latent bound; adding and subtracting
$\psi_\theta(\widehat z_k^\vartheta)$ and applying the decoder regularity bounds
yields the state bound. Continuity of the rollout and stage cost implies
continuity of the finite-horizon objective, and the extreme-value theorem
guarantees a minimizer on compact $C$.
\end{proof}

The Lipschitz bounds use constants defined up to the training horizon $H$,
but the existence of a minimizer needs only continuity, which the rollout has
for any finite horizon. This guarantees the existence of at least one minimizer; it does not guarantee
uniqueness, convexity, or global convergence of a local numerical optimizer.

\subsection{WHY A SHARED DICTIONARY IS NOT ENOUGH}
\label{subsec:theory_projection}
The remaining question is whether one shared dictionary is expressive enough for
the whole closed-loop family. Let $\Phi=(\varphi_1,\dots,\varphi_{d_z})^\top$
with span $\mathcal V_\Phi$. To measure how well a dictionary closes under
$F_\theta$, we need to say on which states the error is counted. We use the
states that the closed loop visits, and let $\mu_\theta$ be the distribution of
$x_k^\theta(x_0)$ when $x_0$ is drawn uniformly from $\mathcal X_0$ and $k$
uniformly from $\{0,\dots,H-1\}$. The
training pairs $(x_k,x_{k+1})$ at parameter $\theta$ are drawn from this distribution, and $\mu_\theta$ depends on $\theta$ because the visited states do. Since the dictionary observables are
continuous and the states lie in the compact set $\mathcal X_H$, they and their
one-step compositions are square integrable under $\mu_\theta$.
For a vector-valued $G=(g_1,\dots,g_{d_z})^\top$ we write
\begin{equation}
  \|G\|_{L^2(\mu_\theta)}^2:=\sum_{j=1}^{d_z}\int_{\mathcal X}g_j(x)^2\,\mathrm d\mu_\theta(x),
  \label{eq:l2_norm_def}
\end{equation}
the usual norm of $L^2(\mu_\theta;\mathbb R^{d_z})$. Here $\mu_\theta$
weights the projection error and is distinct from the fixed reference measure
$\mu$ defining $\mathcal G=L^2(\mathcal X,\mu)$ in
Section~\ref{subsec:koopman_preliminaries}. Following the EDMD projection
viewpoint~\cite{Williams2015EDMD}, the best one-step closure residual available
to \emph{any} choice of latent matrix is
\begin{equation}
  \mathcal R_\theta(\Phi):=\inf_{A\in\mathbb R^{d_z\times d_z}}
  \left\|\Phi\circ F_\theta-A\Phi\right\|_{L^2(\mu_\theta)}.
  \label{eq:closure_risk}
\end{equation}

\begin{proposition}[Fixed-dictionary projection floor]
\label{prop:projection_floor}
Let $P_{\mathcal V_\Phi}^\theta$ be the orthogonal projection onto
$\mathcal V_\Phi$ in $L^2(\mu_\theta)$. Then
\begin{equation}
  \mathcal R_\theta(\Phi)^2=\sum_{j=1}^{d_z}
  \left\|(I-P_{\mathcal V_\Phi}^\theta)(\varphi_j\circ F_\theta)\right\|_{L^2(\mu_\theta)}^2.
  \label{eq:projection_floor}
\end{equation}
\end{proposition}

\begin{proof}
By \eqref{eq:l2_norm_def} the objective in \eqref{eq:closure_risk} separates
across components, and the $j$th component involves only the $j$th row $a_j$ of
$A$:
\begin{equation*}
  \left\|\Phi\circ F_\theta-A\Phi\right\|_{L^2(\mu_\theta)}^2
  =\sum_{j=1}^{d_z}\left\|\varphi_j\circ F_\theta-a_j^\top\Phi\right\|_{L^2(\mu_\theta)}^2 .
\end{equation*}
The rows are unconstrained and independent, so the infimum may be taken
separately in each. As $a_j$ ranges over $\mathbb R^{d_z}$, $a_j^\top\Phi$ ranges
over all of $\mathcal V_\Phi$, and the closest element of a closed subspace to a
given function is its orthogonal projection. Hence the $j$th infimum equals
$\|(I-P_{\mathcal V_\Phi}^\theta)(\varphi_j\circ F_\theta)\|_{L^2(\mu_\theta)}^2$,
and summing gives \eqref{eq:projection_floor}.
\end{proof}

Because the infimum in \eqref{eq:closure_risk} is over \emph{all} matrices $A$, whatever matrix the operator network returns at a given $\theta$ has a one-step residual of at least $\mathcal R_\theta(\Phi)$. The right-hand side of \eqref{eq:projection_floor}
depends only on $\Phi$ and $F_\theta$ and not on how the map
$\theta\mapsto K(\theta)$ is parameterized. Making that map deeper, wider or
otherwise more expressive therefore cannot reduce the part of the closure error
produced by $\varphi_j\circ F_\theta$ lying outside $\mathcal V_\Phi$. The only
way to reduce it is to change $\mathcal V_\Phi$ itself, and
conditioning the encoder and decoder on $\theta$ does this.

Let $\mathfrak D$ be an admissible class of $d_z$-dimensional dictionaries.
We take $\mathfrak D$ to consist of the dictionaries
$\Phi=(x_1,\dots,x_n,g_1,\dots,g_{d_z-n})$ whose first $n$ components are the
state coordinates, with each $g_i$ square integrable. This keeps the state
recoverable from the latent coordinates, and it also fixes the scale of $\Phi$.
The residual is homogeneous, $\mathcal R_\theta(s\Phi)=|s|\,\mathcal R_\theta(\Phi)$, which implies that on a class closed under rescaling both sides of the inequality below would be
zero. The class is a device for the analysis; the learned encoder is not
constrained in this way. Because a shared dictionary must serve every member of the family, the quantity it faces is $\sup_{\theta\in C}\mathcal R_\theta(\Phi)$, minimized over
$\Phi\in\mathfrak D$. A conditioned dictionary may be chosen separately at each
$\theta$, so the quantity it faces is
$\sup_{\theta\in C}\inf_{\Phi\in\mathfrak D}\mathcal R_\theta(\Phi)$. These are
ordered:
\begin{equation}
  \inf_{\Phi\in\mathfrak D}\sup_{\theta\in C}\mathcal R_\theta(\Phi)
  \geq
  \sup_{\theta\in C}\inf_{\Phi\in\mathfrak D}\mathcal R_\theta(\Phi).
  \label{eq:minimax}
\end{equation}
Inequality \eqref{eq:minimax} holds for any function of two variables; taken alone, it therefore only shows that conditioning cannot hurt. It becomes useful once we
know the gap can be strictly positive. The following example shows this, and
the gap there is a property of the closed-loop maps themselves and does not depend on the amount of data or the quality of training.

\begin{remark}[A strict gap]
\label{rem:strict_separation}
Consider the plant $f(x,u)=(ax_1+u_1,\ \delta x_2+u_2)$ on $\mathbb R^2$ with
$a,\delta\neq0$ and the feedback family
$\pi_\theta(x)=(\theta_1\gamma x_2^2,\ \theta_2cx_1^2)$ with $\gamma,c\neq0$,
which has the form \eqref{eq:linear_in_parameter_policy} and satisfies
$\pi_\theta(0)=0$. Take the two controllers $\theta_a=(0,1)$ and
$\theta_b=(1,0)$, for which
\begin{equation*}
  F_{\theta_a}(x)=(ax_1,\ \delta x_2+cx_1^2),
  \qquad
  F_{\theta_b}(x)=(ax_1+\gamma x_2^2,\ \delta x_2).
\end{equation*}
With $n=2$ and $d_z=3$, the dictionaries in $\mathfrak D$ have the form $\Phi=(x_1,x_2,g)$ with $g$ square integrable. We assume that
$\mu_{\theta_a}$ and $\mu_{\theta_b}$ have densities bounded below by some
$m>0$ on a common open set $U$. With $g=x_1^2$ the span is invariant under
$F_{\theta_a}$, and with $g=x_2^2$ it is invariant under $F_{\theta_b}$, so a
conditioned dictionary gives $\mathcal R_\theta=0$ for both controllers.

For a shared dictionary, the residual stays bounded away from zero. Let $\mathcal V$ be its span and
$d(h)$ the $L^2(U)$ distance from $h$ to $\mathcal V$. Since $x_2\in\mathcal V$,
the residual of the component $x_2$ under $F_{\theta_a}$ is at least the
distance from $\delta x_2+cx_1^2$ to $\mathcal V$, so
$\mathcal R_{\theta_a}(\Phi)\geq|c|\sqrt m\,d(x_1^2)$. In the same way,
$\mathcal R_{\theta_b}(\Phi)\geq|\gamma|\sqrt m\,d(x_2^2)$. The space
$W=\operatorname{span}\{x_1,x_2,x_1^2,x_2^2\}$ is four-dimensional in $L^2(U)$
and $\mathcal V$ is at most three-dimensional, and hence $W$ contains a unit vector
$w=\alpha_1x_1+\alpha_2x_2+\beta_1x_1^2+\beta_2x_2^2$ orthogonal to
$\mathcal V$. Because $d$ is a seminorm that vanishes on $x_1$ and $x_2$,
$1=d(w)\leq|\beta_1|\,d(x_1^2)+|\beta_2|\,d(x_2^2)$. The coefficients of unit
vectors in $W$ are bounded by a constant $c_W$ that depends only on $U$, which
gives $\max\{d(x_1^2),d(x_2^2)\}\geq1/(2c_W)$ for every $\Phi\in\mathfrak D$.
Hence, with $C=\{\theta_a,\theta_b\}$, the right-hand side of
\eqref{eq:minimax} is zero, while the left-hand side is at least
$\sqrt m\min\{|c|,|\gamma|\}/(2c_W)>0$. For intermediate parameters the closed
loop couples $x_1$ and $x_2$ in both directions, and neither dictionary is
exact there. The example is only meant to show that the gap in
\eqref{eq:minimax} can be strict.
\end{remark}

Full conditioning therefore enlarges the class of representations available, and
for some families the enlargement is strict. Whether it also improves
finite-data training is a separate, empirical question, which we address
next.

\section{EXPERIMENTS AND RESULTS}
\label{sec:experiments}

The evaluation is organized around three questions. (Q1) Does full policy
conditioning of the coordinates improve prediction relative to conditioning only
the operator, and relative to a bank of per-parameter models? (Q2) Can a single
policy-conditioned model predict closed-loop trajectories across a continuum of
feedback parameters? (Q3) Can the resulting representation serve as a surrogate
for one-shot feedback-parameter selection? Q1 and Q2 evaluate the representation
itself, and Q3 evaluates its use for control. Throughout, the deployment region $C$
is the region returned by the test-and-shrink procedure of Section~\ref{subsec:verification}, with
$\varepsilon_{\max}=15\%$ ($20\%$ for the manipulator), $H=12$ and $N_s=1500$, and the test
parameters of Q2 are fresh samples from it. Fig.~\ref{fig:hull_shrinkage} illustrates
the construction on a representative benchmark.

\subsection{BENCHMARKS AND PROTOCOL}
\label{subsec:benchmarks}

\begin{table}[t]
\centering
\caption{Benchmark systems, feedback families, and model sizes.
All systems use horizon $H=12$. The Network column gives the number of
linear layers $\times$ hidden width, used for the encoder, the decoder and the
operator network alike; for example, $2\times32$ is one hidden layer of $32$
units.}
\label{tab:benchmark_systems}
\footnotesize
\setlength{\tabcolsep}{3.5pt}
\begin{tabular}{lcccc}
\hline
System & $n$ & $p$ & $d_z$ & Network \\
\hline
Van der Pol                   & 2 & 4 & 10 & 2\,$\times$\,32 \\
Polynomial slow manifold      & 2 & 4 & 10 & 2\,$\times$\,32 \\
Two-link planar manipulator   & 4 & 4 & 10 & 2\,$\times$\,32 \\
Cartpole                      & 4 & 4 & 10 & 2\,$\times$\,32 \\
MuJoCo InvertedPendulum       & 4 & 4 & 10 & 2\,$\times$\,32 \\
MuJoCo InvertedDoublePendulum & 6 & 6 & 20 & 3\,$\times$\,128 \\
\hline
\end{tabular}
\end{table}

We evaluate on nonlinear systems of increasing dimension, each written as a
parameterized closed-loop family \eqref{eq:closed_loop_map}
(Table~\ref{tab:benchmark_systems}). The analytic two-dimensional systems test
whether smooth parameter-dependent closed-loop dynamics can be learned in a
controlled setting. The manipulator and cartpole add stronger nonlinearity and
feedback sensitivity. The MuJoCo systems test the framework in a higher-fidelity
simulator whose transition map is not available in closed form.
All models are trained with $\lambda_{\mathrm{rec}}=1$,
$\lambda_{\mathrm{cons}}=0.1$ and $\lambda_{\mathrm{reg}}=10^{-4}$ in
\eqref{eq:total_loss}. The prediction and linearity weights
$(\lambda_{\mathrm{pred}},\lambda_{\mathrm{lin}})$ are $(2,1)$ for VdP,
$(4,1)$ for the slow-manifold system and the manipulator, and $(8,4)$ for
cartpole and the two MuJoCo systems. For each system the test-and-shrink
loop of Section~\ref{subsec:verification} is repeated with the final model
until at least $95\%$ of a fresh batch of $1500$ parameters passes the test,
which required at most $14$ iterations.

\subsubsection{VAN DER POL OSCILLATOR}
The first benchmark is the controlled VdP oscillator with state
$x=[x_1,x_2]^\top$ and dynamics $\dot x_1=x_2$,
$\dot x_2=(1-x_1^2)x_2-x_1+u$. A four-parameter feedback family is used,
\begin{equation}
  u=-\theta_1x_1-\theta_2x_2-\theta_3x_1^3-\theta_4x_1^2x_2,
  \label{eq:vdp_feedback}
\end{equation}
with $\Theta_{\mathrm{VdP}}=[0,5]\times[0,5]\times[-1,1]\times[-1,1]$. Training
trajectories are initialized from $x_0\in[-1,1]^2$ and sampled with
$\Delta t=0.05$\,s. In this low-dimensional benchmark, the feedback
parameter continuously reshapes the transient closed-loop vector field.

\subsubsection{POLYNOMIAL SLOW-MANIFOLD SYSTEM}
\label{subsubsec:slow_manifold}
The second analytic benchmark is a two-dimensional polynomial system with a
nonlinear attracting manifold, $\dot x_1=-0.1x_1+u$ and
$\dot x_2=-(x_2-x_1^2)$. The second state relaxes toward $x_1^2$ while the
control acts through the first. The feedback family and
$\Theta$ are as in \eqref{eq:vdp_feedback}, and trajectories are sampled at
$\Delta t=0.05$\,s. This benchmark is informative for Koopman representation
learning because the dynamics contain an explicit nonlinear manifold and the
feedback family includes nonlinear state basis functions. The useful observable coordinates are therefore themselves nonlinear in the physical state.
It also shows how fragile an exact closure can be. Without feedback the
system closes on the observables $(x_1,x_2,x_1^2)$, since
$\tfrac{\mathrm d}{\mathrm dt}x_1^2=-0.2x_1^2$ stays in their span. The linear
term $-\theta_2x_2$ of the feedback alone gives
$\tfrac{\mathrm d}{\mathrm dt}x_1^2=-0.2x_1^2-2\theta_2x_1x_2$, the derivative
of $x_1x_2$ then brings in $x_1^3$ and $x_2^2$, and each new observable adds
monomials of higher degree, so no finite set of monomials closes.

\subsubsection{TWO-LINK PLANAR MANIPULATOR}
The third benchmark is a fully actuated two-link planar manipulator with state
$x=[q_1,q_2,\dot q_1,\dot q_2]^\top$ and nonlinear dynamics written in
manipulator form as $M(q)\ddot q+h(q,\dot q)=\tau$, where $q=[q_1,q_2]^\top$ and
$\tau=[\tau_1,\tau_2]^\top$. The experiments use $m_1=m_2=1$, $l_1=l_2=1$, and
$g=9.81$. A decentralized proportional-derivative feedback family is used,
$\tau_i=-(k_{p,i}q_i+k_{d,i}\dot q_i)$ for $i=1,2$, with
$\theta=[k_{p,1},k_{d,1},k_{p,2},k_{d,2}]^\top$ and
$\Theta_{\mathrm{man}}=[0,40]\times[0,20]\times[0,20]\times[0,20]$. Both torques
are saturated to $[-30,30]$ and the continuous dynamics are sampled at
$\Delta t=0.05$\,s. Unlike the preceding two-dimensional systems, this benchmark
is four-dimensional and strongly coupled, so it tests whether
the same architecture remains effective when one parameter vector
changes two physical feedback loops. It is also the only benchmark whose
verified region uses $\varepsilon_{\max}=20\%$, because the test-and-shrink loop does not
converge at $15\%$ on this system.

\subsubsection{CARTPOLE}
The cartpole is an underactuated four-dimensional mechanical system with state
$x=[p,\dot p,\varphi,\dot\varphi]^\top$, where $p$ is the cart displacement and
$\varphi$ the pole angle from upright, and physical parameters $M=1$, $m=0.1$,
$l=0.5$, $g=9.81$. The feedback law is
\begin{equation}
  u=-\left(\theta_1p+\theta_2\dot p+\theta_3\sin\varphi+\theta_4\dot\varphi\right).
  \label{eq:cartpole_feedback}
\end{equation}
Training initial conditions are sampled from $p,\dot p\in[-0.5,0.5]$,
$\varphi\in[-0.3,0.3]$, $\dot\varphi\in[-0.8,0.8]$, with
$\Delta t=0.05$\,s. This benchmark tests the representation on an
underactuated unstable system for which changes in the feedback parameters
substantially alter both the transient response and long-horizon stability; it
is also the system used for the architectural ablations.

\subsubsection{MUJOCO ENVIRONMENTS}
To evaluate the method beyond explicitly coded analytic dynamics, we use the
Gymnasium MuJoCo \texttt{InvertedPendulum-v5} and
\texttt{InvertedDoublePendulum-v5} environments. For the inverted pendulum the
four-dimensional state $x=[p,\dot p,\varphi,\dot\varphi]^\top$ is obtained by
reordering the MuJoCo observation, and the feedback action takes the form
\eqref{eq:cartpole_feedback}, clipped to the action limits imposed by the
environment. The double pendulum uses the analogous six-parameter family on its
six-dimensional state. Initial conditions match the cartpole ranges and
simulator transitions are sampled at the environment step, $\Delta t=0.04$\,s for
the inverted pendulum and $0.05$\,s for the double pendulum. Unlike the analytic
cartpole, the transition map used for training and evaluation is generated
directly by the simulator, without an explicit plant model.

\noindent\textbf{Metrics.} For each test pair $(x_0,\theta)$ we report the
maximum normalized multi-step error
\begin{equation}
  e(x_0,\theta)=\max_{1\leq k\leq H}
  \frac{\|\hat x_k^\theta-F_\theta^k(x_0)\|_2}
       {\max\!\left(\|F_\theta^k(x_0)\|_2,0.1\right)}\times100,
  \label{eq:prediction_metric}
\end{equation}
summarized in Table~\ref{tab:prediction_results} by the mean and standard deviation; Fig.~\ref{fig:prediction_results} also marks the median. The maximum over the horizon penalizes the
worst error during the rollout, and the denominator floor prevents artificially
large relative errors near the equilibrium. For the control application, we
report whether the controller drives the system to the reference point and
how much online computation it needs. We use this success criterion because a weighted closed-loop cost would make the comparison depend on a particular choice of $Q_x$ and $R_u$. Success and runtime are always read
together, since a controller that terminates early
accumulates less computation.

\noindent\textbf{Baselines.} We compare against PKNN~\cite{Guo_parametric_Koopman},
Deep Koopman, EDMD, and Kernel-EDMD for prediction. PKNN learns one
shared dictionary and conditions only the operator on $\theta$. Deep Koopman is
trained on the same mixed-parameter trajectories as our model but does not
receive $\theta$, so it fits one linear latent model to the whole family. EDMD
uses a dictionary of monomials (with sines and cosines of the angles for the
manipulator and cartpole), and Kernel-EDMD uses Gaussian radial basis
features. Both are fitted at one parameter, the centroid of the deployment region,
and applied unchanged at the test parameters. The last three baselines
therefore show what is lost when the model ignores $\theta$. For the control
example we use nonlinear MPC (NMPC) and receding-horizon Koopman MPC (K-MPC) built on the Deep Koopman and EDMD models as
reference points. All methods use the hyperparameters of their original sources and are evaluated from identical initial
conditions on the same true system, with matched actuator bounds, termination
conditions, integration step, episode length, and cost weights in the physical coordinates.

\subsection{Q1: WHY FULL POLICY CONDITIONING}
\label{subsec:results_ablation}

\begin{figure}[t]
    \centering
    \includegraphics[width=\columnwidth]{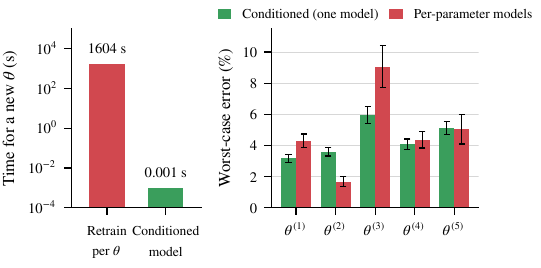}
    \caption{A single policy-conditioned model versus separately trained
parameter-specific Koopman models on cartpole. Left: cost of obtaining a
predictor for a new feedback parameter. Right: mean worst-case
prediction error over $H=12$ steps at five sampled parameter vectors
$\theta^{(1)},\dots,\theta^{(5)}$.}
    \label{fig:ablation1}
\end{figure}

The ablations follow the progression from per-parameter models, to an
operator-conditioned model with a fixed dictionary, to the fully conditioned
representation. They are run on cartpole, where the feedback parameter strongly
affects the closed-loop dynamics, and Ablation~B is repeated on the other
analytic benchmarks. All runs use five random seeds with matched dynamics,
controller parameterization, horizon $H=12$, latent dimension $d_z=10$, state
normalization, optimizer, objective, and data budget.

\noindent\textbf{Ablation A: conditioning versus per-parameter retraining.}
The conditioned model \eqref{eq:fully_conditioned} is trained once on a random
sweep of feedback parameters; the baseline \eqref{eq:unconditioned} has no access
to $\theta$ and is therefore retrained separately at each of five tested parameter vectors.
Fig.~\ref{fig:ablation1} reports the cost of obtaining a predictor for a new
parameter and the resulting accuracy. One parameter-specific model requires
approximately $1.6\times10^3$\,s ($\approx26.7$\,min) of training,
and this cost recurs for every new controller parameter, whereas the conditioned
model is trained once over a parameter distribution and evaluated at a
new $\theta$ without retraining. Accuracy does not degrade appreciably, and the
conditioned model is more accurate at some of the displayed parameters.
A parameter-specific model is specialized to one closed-loop system, but
specialization does not guarantee lower held-out error at every parameter,
because optimization, finite data, and initialization also matter.

\begin{figure}[t]
    \centering
    \includegraphics[width=\columnwidth]{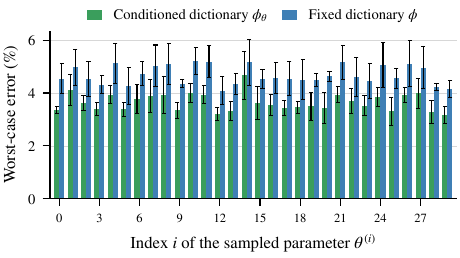}
    \caption{Mean worst-case multi-step prediction error at $30$
feedback parameters $\theta^{(0)},\dots,\theta^{(29)}$ drawn uniformly from the
cartpole hull built in Phases~1 and~2, with $K(\theta)$ conditioned in both variants, averaged over five seeds.
Conditioning the observable coordinates lowers the error at all $30$
parameters.}
    \label{fig:ablation2_accuracy}
\end{figure}

\noindent\textbf{Ablation B: fixed versus conditioned dictionary.}
Both variants use a parameter-conditioned operator network, which gives each of them access to $K(\theta)$; the only difference is whether the encoder and decoder also
receive $\theta$, i.e.\ \eqref{eq:fully_conditioned} versus
\eqref{eq:operator_conditioned}. They share the same random-parameter dataset,
loss, optimizer settings, horizon, epochs, and sample budget, and any difference can thus be attributed to dictionary conditioning alone. The shared
dataset has $20000$ cartpole trajectories with parameters drawn uniformly from $\Theta$. Both variants are tested at $30$ parameters sampled uniformly
from the hull built in Phases~1 and~2, each from the $81$ initial conditions of
$\mathcal X_{\mathrm{ver}}$. Over these parameters
(Fig.~\ref{fig:ablation2_accuracy}) the fully conditioned model achieves a lower mean
worst-case error at every parameter. Averaged over all $30$, the
error falls from $4.68\%$ to $3.65\%$, about $22\%$ lower. The conditioned
model also varies less across seeds (standard deviation $0.42\%$ against
$0.57\%$), so on cartpole conditioning improves both mean accuracy and consistency across initializations.
On the other analytic benchmarks the differences are smaller. Conditioning
is slightly better on the slow-manifold system and ties on the manipulator,
while on VdP the fixed dictionary is more accurate ($7.32\%$ against $8.06\%$). These values differ from Table~\ref{tab:prediction_results} because the ablation has its own models and test set.

Together the ablations support the two design decisions behind the architecture. One conditioned model removes per-parameter retraining without an appreciable loss in accuracy (Ablation~A), and conditioning the encoder and decoder lowers the error on cartpole, where the feedback parameter strongly changes the closed loop (Ablation~B).

\subsection{Q2: CLOSED-LOOP PREDICTION ACROSS FEEDBACK PARAMETERS}
\label{subsec:results_prediction}

\begin{table*}[t]
\centering
\caption{Closed-loop prediction error (\%): mean $\pm$ standard deviation of the
maximum normalized multi-step error \eqref{eq:prediction_metric} over
$N_\theta=30$ feedback parameters and $N_{\mathrm{IC}}=20$ initial conditions
per parameter, $H=12$. The parameters are fresh samples from the
verified regions of Section~\ref{subsec:benchmarks}, drawn with a seed not used in
training. IP and IDP: MuJoCo inverted and inverted double pendulum.}
\label{tab:prediction_results}
\begin{tabular}{lcccccc}
\hline
Method & Van der Pol & Slow manifold & Manipulator & Cartpole & MuJoCo IP & MuJoCo IDP \\
\hline
Ours                               & $3.57\pm2.41$   & $2.38\pm1.63$   & $\bm{5.61\pm2.02}$ & $\bm{3.89\pm1.58}$ & $\bm{7.76\pm3.39}$ & $\bm{4.32\pm2.01}$ \\
PKNN~\cite{Guo_parametric_Koopman} & $\bm{0.63\pm0.53}$ & $\bm{1.11\pm0.86}$ & $6.09\pm3.72$ & $134.58\pm126.99$ & $8.04\pm4.13$ & $87.65\pm35.53$ \\
EDMD~\cite{Williams2015EDMD}       & $21.38\pm14.94$ & $11.29\pm8.32$ & $9.20\pm5.94$     & $7.55\pm4.51$ & $58.10\pm39.95$ & $16.00\pm10.97$ \\
Deep Koopman~\cite{Lusch2018}      & $31.31\pm24.89$ & $32.30\pm21.85$ & $85.97\pm22.40$ & $90.60\pm24.20$ & $216.38\pm103.04$ & $514.57\pm266.73$ \\
Kernel-EDMD~\cite{williams2015kernel} & $23.79\pm14.72$ & $18.80\pm9.36$ & $24.49\pm13.82$ & $17.19\pm11.06$ & $61.20\pm34.15$ & $97.16\pm44.69$ \\
\hline
\end{tabular}
\end{table*}

\begin{figure*}[t]
\centering
\includegraphics[width=\linewidth]{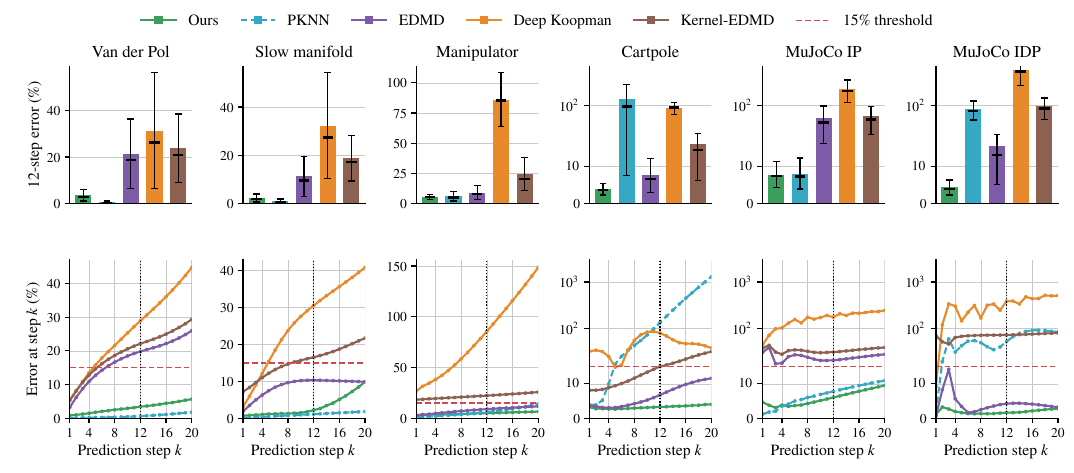}
\caption{Closed-loop prediction error on six nonlinear benchmarks. Top:
worst-case relative error over $H=12$ steps. Bottom: worst-case relative error
versus prediction step, extended to $k=20$ beyond the training horizon. Same test cases as Table~\ref{tab:prediction_results}.}
\label{fig:prediction_results}
\end{figure*}


We next test whether one learned representation predicts the trajectories of a
family of closed-loop systems as the feedback parameter varies. For each
benchmark we draw $N_\theta=30$ fresh parameters uniformly from $C$, with a seed not used in training, and
$N_{\mathrm{IC}}=20$ initial conditions per parameter. We simulate the true closed
loop for $H=12$ steps and generate the prediction entirely through the learned
representation \eqref{eq:latent_rollout}, with results in
Table~\ref{tab:prediction_results} and Fig.~\ref{fig:prediction_results}.

Across the six benchmarks the proposed model is the most consistent. Its
mean $12$-step errors remain below the $15\%$ threshold on every system, between
$2.4\%$ and $7.8\%$. This error grows slowly with the
horizon and stays below the threshold out to $k=20$ on all six systems, beyond the horizon used in training. PKNN is very accurate on VdP and the slow-manifold system, with mean errors of $0.6\%$ and $1.1\%$, and stays
below the threshold on the manipulator and the inverted pendulum. On
cartpole and the double pendulum, however, its errors grow rapidly with the horizon and
exceed the threshold, on cartpole by almost an order of magnitude. Parameterizing only the latent operator can therefore be
highly effective for some families but is not uniformly reliable across the tested ones. Deep Koopman and
Kernel-EDMD exceed the threshold on all six systems. EDMD, fitted at the centroid
of $C$, stays below it on the slow-manifold system, the manipulator and
cartpole, which suggests that the closed loop varies little across the verified regions of these systems.

The advantage of the proposed model over PKNN appears on the mechanical systems and is
largest on cartpole and the double pendulum. On the two-dimensional benchmarks PKNN attains
the lowest errors while the proposed method stays below the threshold; on the
four-dimensional manipulator and cartpole the proposed model attains the lowest
errors and the most reliable long-horizon propagation. On the MuJoCo inverted
pendulum PKNN comes close ($8.0\%$ against $7.8\%$), and on the double pendulum
our model is the only one below the threshold. This pattern is in line with
the motivation behind Proposition~\ref{prop:projection_floor}. Conditioning
the encoder and decoder should matter most when the feedback parameter changes
the closed-loop transient and hence the observable coordinates required for
accurate prediction. The comparison with PKNN does not isolate this effect on
its own, and Ablation~B is the
controlled test of it.

\subsection{Q3: ONE-SHOT FEEDBACK-PARAMETER SELECTION}
\label{subsec:results_one_shot}

\begin{figure*}[t]
    \centering
    \includegraphics[width=\linewidth]{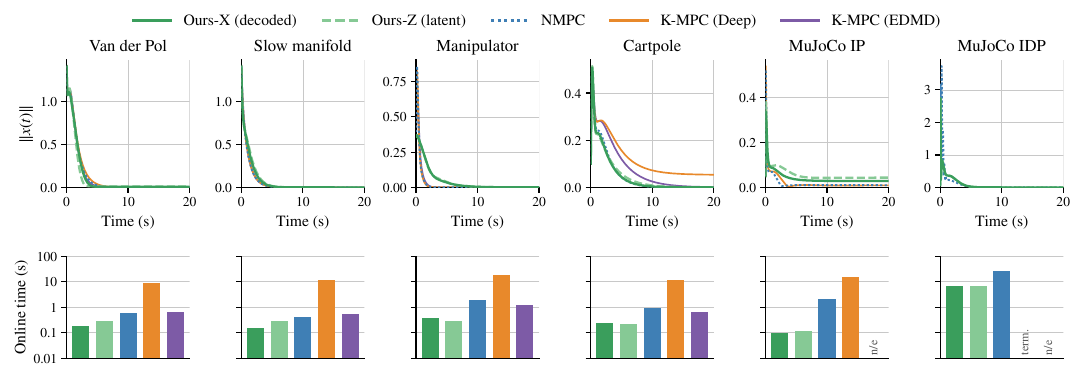}
    \caption{One-shot feedback-parameter selection. Top: convergence of
the states under the selected controller across the benchmarks. Bottom:
online computational cost per episode in seconds (Table~\ref{tab:one_shot_summary}), including the one-time selection;
``term.'': terminated, ``n/e'': not evaluated. Both variants select parameters that
drive every benchmark to the equilibrium, at the lowest online cost among the
compared methods.}
    \label{fig:one_shot}
\end{figure*}

\begin{table*}[t]
\centering
\caption{One-shot feedback-parameter selection compared with reference controllers,
per benchmark. A checkmark under ``Succ.'' means the controller
stabilized the system in all five seeds, and $t_{\mathrm{tot}}$ is the total online computation
time per episode, averaged over five seeds. For Ours-Z and Ours-X it
includes the one-time selection. Ours uses the verified regions of
Section~\ref{subsec:benchmarks}.}
\label{tab:one_shot_summary}

\setlength{\tabcolsep}{5pt}
\renewcommand{\arraystretch}{1.12}

\resizebox{\textwidth}{!}{%
\begin{tabular}{l
                cc
                cc
                cc
                cc
                cc}
\toprule

& \multicolumn{2}{c}{\textbf{Ours-Z}}
& \multicolumn{2}{c}{\textbf{Ours-X}}
& \multicolumn{2}{c}{\textbf{NMPC}}
& \multicolumn{2}{c}{\textbf{K-MPC (Deep)}}
& \multicolumn{2}{c}{\textbf{K-MPC (EDMD)}} \\

\cmidrule(lr){2-3}
\cmidrule(lr){4-5}
\cmidrule(lr){6-7}
\cmidrule(lr){8-9}
\cmidrule(lr){10-11}

\textbf{System}
& \textbf{Succ.} & $\boldsymbol{t_{\rm tot}}$
& \textbf{Succ.} & $\boldsymbol{t_{\rm tot}}$
& \textbf{Succ.} & $\boldsymbol{t_{\rm tot}}$
& \textbf{Succ.} & $\boldsymbol{t_{\rm tot}}$
& \textbf{Succ.} & $\boldsymbol{t_{\rm tot}}$ \\

\midrule

Van der Pol
& $\checkmark$ & 0.299
& $\checkmark$ & \textbf{0.175}
& $\checkmark$ & 0.585
& $\checkmark$ & 8.913
& $\checkmark$ & 0.665 \\

Polynomial slow manifold
& $\checkmark$ & 0.297
& $\checkmark$ & \textbf{0.150}
& $\checkmark$ & 0.422
& $\checkmark$ & 11.275
& $\checkmark$ & 0.529 \\

Two-link planar manipulator
& $\checkmark$ & \textbf{0.276}
& $\checkmark$ & 0.390
& $\checkmark$ & 1.855
& $\checkmark$ & 19.007
& $\checkmark$ & 1.202 \\

Cartpole
& $\checkmark$ & \textbf{0.220}
& $\checkmark$ & 0.244
& $\checkmark$ & 0.913
& $\checkmark$ & 12.168
& $\checkmark$ & 0.663 \\

MuJoCo InvertedPendulum
& $\checkmark$ & 0.116
& $\checkmark$ & \textbf{0.100}
& $\checkmark$ & 2.106
& $\checkmark$ & 14.705
& N/E & -- \\

MuJoCo InvertedDoublePendulum
& $\checkmark$ & \textbf{6.507}
& $\checkmark$ & 6.882
& $\checkmark$ & 25.631
& Term. & --
& N/E & -- \\

\bottomrule
\end{tabular}%
}

\vspace{1mm}
\begin{flushleft}
\footnotesize
All computation times are in seconds.
N/E: not evaluated.
Term.: prematurely terminated.
\end{flushleft}

\end{table*}

Finally, we illustrate one use of the representation by selecting the
feedback parameter once, at the initial state.
Ours-Z minimizes the accumulated magnitude of the predicted latent trajectory
\eqref{eq:latent_objective}; Ours-X evaluates the same Koopman-predicted
trajectory after decoding \eqref{eq:decoded_objective}. Both use the same
trained model, verified region, selection horizon, optimizer initialization,
parameter constraints, and tolerances, so their comparison isolates the effect
of the scoring coordinates. Ours-Z and Ours-X perform one initial optimization,
whereas NMPC and receding-horizon Koopman MPC re-solve after every state
transition. The relevant comparison is therefore the total online computation per episode.

Fig.~\ref{fig:one_shot} shows representative true-system trajectories and
Table~\ref{tab:one_shot_summary} the per-benchmark success and online
runtimes. Ours-Z and Ours-X
produce closely aligned closed-loop behavior, nearly indistinguishable at the plotting scale on the four analytic benchmarks, and both stabilize the displayed MuJoCo initial conditions.
The two scoring criteria are not mathematically
equivalent, and the competitive behavior of Ours-Z indicates that the learned latent
coordinates carry information relevant to closed-loop regulation. We regard this as an empirical property of the representation over the verified region and do not claim a general equivalence between latent and physical costs.

NMPC is the strongest adaptive baseline, since it repeatedly optimizes a future
input sequence using the nonlinear dynamics, and it also stabilizes every
benchmark. It requires $1.4$ to $21$ times more online computation than the
selected controllers, however. Both proposed variants
instead solve one low-dimensional optimization at the initial state and then
apply a fixed physical feedback law. The runtimes of Ours-Z and Ours-X are
of the same order, and neither variant is consistently faster, which
indicates that decoder evaluation does not dominate the
present implementation. The computational advantage comes from the one-shot selection itself, and scoring in latent coordinates gives no consistent saving.

\subsection{DISCUSSION AND LIMITATIONS}
\label{subsec:limitations}

Conditioning the coordinates on
$\theta$ enlarges the representation class (Proposition~\ref{prop:projection_floor}
and \eqref{eq:minimax}), and Ablation~B shows that this lowers the error
on cartpole. The benefit is not uniform, however, since on the simpler benchmarks PKNN and the fixed dictionary remain competitive or better. This pattern is consistent with the projection argument, because when $\theta$ has only a mild effect on the transient geometry, there is little closure error for conditioning to remove.

The first limitation is that the verified region $C$ is an empirical construction. Phases~1--3 report the region on which held-out prediction and
admissibility tests were passed, but they do not deliver a probabilistic
statement about untested parameters. Such a statement would require a sampling argument
matched to the construction of the decision set in the manner
of~\cite{campi2018scenario}. The convex-hull representation compounds this, because
convexity is imposed rather than established. Hull-volume estimation also becomes less accurate and needs more samples as the feedback-parameter dimension $p$ grows, and we test only up to $p=6$. The procedure is also expensive, because every
iteration simulates the true closed loop for $1500$ parameters from every
initial condition in $\mathcal X_{\mathrm{ver}}$, and the double pendulum
needed $14$ iterations.

The theory is also conditional, since the constants \eqref{eq:rec_defect}--\eqref{eq:lpsi_def} exist but the analysis does not
control their size, and Theorem~\ref{thm:prediction} accordingly describes how errors propagate without bounding their magnitude. None of the results says the deployed
controller stabilizes the plant; the reported stabilization is empirical, and
SLSQP, started at the mean of the verified samples, returns a local solution.

The controller-selection experiments do not compare closed-loop performance
with controllers tuned directly on the plant. The analysis of
Section~\ref{sec:theory} also does not bound the gap between the selected parameter
and the minimizer of the true cost.

Finally, the one-shot formulation itself assumes the initial state carries enough
information to choose a parameter for the whole episode. This is reasonable for
regulation to a known equilibrium from a bounded initial set, which is the
setting studied here. It is not appropriate when the operating condition
changes mid-episode or when disturbances move the state far outside the region
over which the surrogate was verified. In such cases the natural extension is
periodic re-selection, which would recover part of the adaptivity of a
receding-horizon scheme while retaining the low-dimensional decision variable.

\section{CONCLUSION}
\label{sec:conclusion}
This paper presented a policy-conditioned Koopman representation that learns the
observable coordinates and latent dynamics of a family of nonlinear closed-loop
systems. Conditioning the encoder, operator, and decoder on the feedback
parameter enables prediction across controllers without per-controller
training, and a projection argument shows that a shared dictionary can be strictly less expressive. Experiments show that the fully conditioned model
gives the most reliable prediction on the tested mechanical systems, with
the clearest improvement over a fixed dictionary on cartpole in a matched
comparison. A shared
dictionary remains competitive or better on the simpler benchmarks. As an example
application, the same representation supports one-shot feedback-parameter selection, after which
deployment requires only evaluation of the selected state-feedback law.
Future work includes a certified deployment region
that scales better in $p$ (for example a conformal or Gaussian-process
construction), data-dependent versions of the bounds in Section~\ref{sec:theory}, and experiments on
physical hardware.

\bibliographystyle{IEEEtran}
\bibliography{references}

@article{Lusch2018,
  author  = {Lusch, Bethany and Kutz, J. Nathan and Brunton, Steven L.},
  title   = {Deep Learning for Universal Linear Embeddings of Nonlinear Dynamics},
  journal = {Nature Communications},
  volume  = {9},
  number  = {1},
  pages   = {4950},
  year    = {2018},
  month   = nov,
  doi     = {10.1038/s41467-018-07210-0},
  url     = {https://doi.org/10.1038/s41467-018-07210-0},
  issn    = {2041-1723}
}

@article{Guo_parametric_Koopman,
  author  = {Guo, Yue and Korda, Milan and Kevrekidis, Ioannis G. and Li, Qianxiao},
  title   = {Learning Parametric {Koopman} Decompositions for Prediction and Control},
  journal = {SIAM Journal on Applied Dynamical Systems},
  volume  = {24},
  number  = {1},
  pages   = {744--781},
  year    = {2025},
  doi     = {10.1137/23M1604576},
  url     = {https://doi.org/10.1137/23M1604576}
}

@inproceedings{azencot20a_ConsistentKoopman,
  author    = {Azencot, Omri and Erichson, N. Benjamin and Lin, Vanessa and Mahoney, Michael W.},
  title     = {Forecasting Sequential Data Using Consistent {Koopman} Autoencoders},
  booktitle = {Proceedings of the 37th International Conference on Machine Learning},
  editor    = {Daum\'e III, Hal and Singh, Aarti},
  series    = {Proceedings of Machine Learning Research},
  volume    = {119},
  pages     = {475--485},
  year      = {2020},
  month     = jul,
  publisher = {PMLR},
  url       = {https://proceedings.mlr.press/v119/azencot20a.html}
}

@article{Delauney_convex_1,
  author   = {Barber, C. Bradford and Dobkin, David P. and Huhdanpaa, Hannu},
  title    = {The Quickhull Algorithm for Convex Hulls},
  journal  = {ACM Transactions on Mathematical Software},
  volume   = {22},
  number   = {4},
  pages    = {469--483},
  year     = {1996},
  month    = dec,
  doi      = {10.1145/235815.235821},
  url      = {https://doi.org/10.1145/235815.235821},
  issn     = {0098-3500}
}

@article{Bowyer1981ComputingDT_Delauney,
  title={Computing {Dirichlet} Tessellations},
  author={Adrian Bowyer},
  journal={The Computer Journal},
  year={1981},
  volume={24},
  number={2},
  pages={162--166},
  doi={10.1093/comjnl/24.2.162}
}

@article{Koopman1931,
  author  = {Koopman, B. O.},
  title   = {Hamiltonian Systems and Transformation in {Hilbert} Space},
  journal = {Proceedings of the National Academy of Sciences of the United States of America},
  volume  = {17},
  number  = {5},
  pages   = {315--318},
  year    = {1931},
  doi     = {10.1073/pnas.17.5.315}
}

@article{Brunton2022ModernKoopman,
  author  = {Brunton, Steven L. and Budi\v{s}i\'c, Marko and Kaiser, Eurika and Kutz, J. Nathan},
  title   = {Modern {Koopman} Theory for Dynamical Systems},
  journal = {SIAM Review},
  volume  = {64},
  number  = {2},
  pages   = {229--340},
  year    = {2022},
  doi     = {10.1137/21M1401243}
}

@article{Williams2015EDMD,
  author  = {Williams, Matthew O. and Kevrekidis, Ioannis G. and Rowley, Clarence W.},
  title   = {A Data-Driven Approximation of the {Koopman} Operator: Extending Dynamic Mode Decomposition},
  journal = {Journal of Nonlinear Science},
  volume  = {25},
  number  = {6},
  pages   = {1307--1346},
  year    = {2015},
  doi     = {10.1007/s00332-015-9258-5}
}

@article{Brunton2016InvariantSubspaces,
  author  = {Brunton, Steven L. and Brunton, Bingni W. and Proctor, Joshua L. and Kutz, J. Nathan},
  title   = {Koopman Invariant Subspaces and Finite Linear Representations of Nonlinear Dynamical Systems for Control},
  journal = {PLOS ONE},
  volume  = {11},
  number  = {2},
  pages   = {e0150171},
  year    = {2016},
  doi     = {10.1371/journal.pone.0150171}
}

@article{Li2017EDMDDL,
  author  = {Li, Qianxiao and Dietrich, Felix and Bollt, Erik M. and Kevrekidis, Ioannis G.},
  title   = {Extended Dynamic Mode Decomposition with Dictionary Learning: A Data-Driven Adaptive Spectral Decomposition of the {Koopman} Operator},
  journal = {Chaos},
  volume  = {27},
  number  = {10},
  pages   = {103111},
  year    = {2017},
  doi     = {10.1063/1.4993854}
}

@article{Otto2019LRAE,
  author  = {Otto, Samuel E. and Rowley, Clarence W.},
  title   = {Linearly Recurrent Autoencoder Networks for Learning Dynamics},
  journal = {SIAM Journal on Applied Dynamical Systems},
  volume  = {18},
  number  = {1},
  pages   = {558--593},
  year    = {2019},
  doi     = {10.1137/18M1177846}
}

@article{Proctor2018KoopmanControl,
  author  = {Proctor, Joshua L. and Brunton, Steven L. and Kutz, J. Nathan},
  title   = {Generalizing {Koopman} Theory to Allow for Inputs and Control},
  journal = {SIAM Journal on Applied Dynamical Systems},
  volume  = {17},
  number  = {1},
  pages   = {909--930},
  year    = {2018},
  doi     = {10.1137/16M1062296}
}

@article{Korda2018KoopmanMPC,
  author  = {Korda, Milan and Mezi\'c, Igor},
  title   = {Linear Predictors for Nonlinear Dynamical Systems: {Koopman} Operator Meets Model Predictive Control},
  journal = {Automatica},
  volume  = {93},
  pages   = {149--160},
  year    = {2018},
  doi     = {10.1016/j.automatica.2018.03.046}
}

@article{mezic2005spectral,
  author  = {Mezi{\'c}, Igor},
  title   = {Spectral Properties of Dynamical Systems, Model Reduction and Decompositions},
  journal = {Nonlinear Dynamics},
  year    = {2005},
  volume  = {41},
  pages   = {309--325},
  doi     = {10.1007/s11071-005-2824-x}
}

@article{rowley2009spectral,
  author  = {Rowley, Clarence W. and Mezi{\'c}, Igor and Bagheri, Shervin and Schlatter, Philipp and Henningson, Dan S.},
  title   = {Spectral Analysis of Nonlinear Flows},
  journal = {Journal of Fluid Mechanics},
  year    = {2009},
  volume  = {641},
  pages   = {115--127},
  doi     = {10.1017/S0022112009992059}
}

@article{schmid2010dmd,
  author  = {Schmid, Peter J.},
  title   = {Dynamic Mode Decomposition of Numerical and Experimental Data},
  journal = {Journal of Fluid Mechanics},
  year    = {2010},
  volume  = {656},
  pages   = {5--28},
  doi     = {10.1017/S0022112010001217}
}

@article{williams2015kernel,
  author  = {Williams, Matthew O. and Rowley, Clarence W. and Kevrekidis, Ioannis G.},
  title   = {A Kernel-Based Method for Data-Driven {Koopman} Spectral Analysis},
  journal = {Journal of Computational Dynamics},
  year    = {2015},
  volume  = {2},
  number  = {2},
  pages   = {247--265},
  doi     = {10.3934/jcd.2015005}
}

@inproceedings{takeishi2017learning,
  author    = {Takeishi, Naoya and Kawahara, Yoshinobu and Yairi, Takehisa},
  title     = {Learning {Koopman} Invariant Subspaces for Dynamic Mode Decomposition},
  booktitle = {Advances in Neural Information Processing Systems},
  year      = {2017},
  volume    = {30},
  pages     = {1130--1140}
}

@inproceedings{yeung2019learning,
  author    = {Yeung, Enoch and Kundu, Soumya and Hodas, Nathan},
  title     = {Learning Deep Neural Network Representations for {Koopman} Operators of Nonlinear Dynamical Systems},
  booktitle = {Proceedings of the 2019 American Control Conference},
  year      = {2019},
  pages     = {4832--4839},
  doi       = {10.23919/ACC.2019.8815339}
}

@inproceedings{fathi2024course,
  author    = {Fathi, Mahan and Gehring, Clement and Pilault, Jonathan and Kanaa, David and Bacon, Pierre-Luc and Goroshin, Ross},
  title     = {Course Correcting {Koopman} Representations},
  booktitle = {International Conference on Learning Representations},
  year      = {2024}
}

@article{proctor2016dmdc,
  author  = {Proctor, Joshua L. and Brunton, Steven L. and Kutz, J. Nathan},
  title   = {Dynamic Mode Decomposition with Control},
  journal = {SIAM Journal on Applied Dynamical Systems},
  year    = {2016},
  volume  = {15},
  number  = {1},
  pages   = {142--161},
  doi     = {10.1137/15M1013857}
}

@article{cibulka2025dictionaryfree,
  author  = {Cibulka, V{\'i}t and Korda, Milan and Hani{\v{s}}, Tom{\'a}{\v{s}}},
  title   = {Dictionary-Free {Koopman} Model Predictive Control with Nonlinear Input Transformation},
  journal = {SIAM Journal on Applied Dynamical Systems},
  year    = {2025},
  volume  = {24},
  number  = {1},
  pages   = {94--130},
  doi     = {10.1137/23M1597174}
}

@article{haseli2026kcf,
  author  = {Haseli, Masih and Cort{\'e}s, Jorge},
  title   = {Modeling Nonlinear Control Systems via {Koopman Control Family}: Universal Forms and Subspace Invariance Proximity},
  journal = {Automatica},
  year    = {2026},
  volume  = {185},
  pages   = {112722},
  doi     = {10.1016/j.automatica.2025.112722}
}

@article{andreuzzi2023parametric,
  author  = {Andreuzzi, Francesco and Demo, Nicola and Rozza, Gianluigi},
  title   = {A Dynamic Mode Decomposition Extension for the Forecasting of Parametric Dynamical Systems},
  journal = {SIAM Journal on Applied Dynamical Systems},
  year    = {2023},
  volume  = {22},
  number  = {3},
  pages   = {2432--2458},
  doi     = {10.1137/22M1481658}
}

@article{huhn2023parametric,
  author  = {Huhn, Quincy A. and Tano, Mauricio E. and Ragusa, Jean C. and Choi, Youngsoo},
  title   = {Parametric Dynamic Mode Decomposition for Reduced Order Modeling},
  journal = {Journal of Computational Physics},
  year    = {2023},
  volume  = {475},
  pages   = {111852},
  doi     = {10.1016/j.jcp.2022.111852}
}

@inproceedings{sui2015safeopt,
  author    = {Sui, Yanan and Gotovos, Alkis and Burdick, Joel and Krause, Andreas},
  title     = {Safe Exploration for Optimization with {Gaussian} Processes},
  booktitle = {Proceedings of the 32nd International Conference on Machine Learning},
  series    = {Proceedings of Machine Learning Research},
  volume    = {37},
  pages     = {997--1005},
  year      = {2015},
  publisher = {PMLR}
}

@inproceedings{dorschel2021safe,
  author    = {D{\"o}rschel, Lorenz and Stenger, David and Abel, Dirk},
  title     = {Safe {Bayesian} Optimisation for Controller Design by Utilising the Parameter Space Approach},
  booktitle = {Proceedings of the 3rd Conference on Learning for Dynamics and Control},
  series    = {Proceedings of Machine Learning Research},
  volume    = {144},
  pages     = {299--311},
  year      = {2021},
  publisher = {PMLR}
}

@inproceedings{widmer2023tuning,
  author    = {Widmer, Daniel and Kang, Dongho and Sukhija, Bhavya and H{\"u}botter, Jonas and Krause, Andreas and Coros, Stelian},
  title     = {Tuning Legged Locomotion Controllers via Safe {Bayesian} Optimization},
  booktitle = {Proceedings of the 7th Conference on Robot Learning},
  series    = {Proceedings of Machine Learning Research},
  volume    = {229},
  pages     = {2444--2464},
  year      = {2023},
  publisher = {PMLR}
}

@inproceedings{wei2025highdimensional,
  author    = {Wei, Yunyue and Yi, Zeji and Li, Hongda and Soedarmadji, Saraswati and Sui, Yanan},
  title     = {Safe {Bayesian} Optimization for the Control of High-Dimensional Embodied Systems},
  booktitle = {Proceedings of the 8th Conference on Robot Learning},
  series    = {Proceedings of Machine Learning Research},
  volume    = {270},
  pages     = {4771--4792},
  year      = {2025},
  publisher = {PMLR}
}

@book{campi2018scenario,
  author    = {Campi, Marco C. and Garatti, Simone},
  title     = {Introduction to the Scenario Approach},
  publisher = {Society for Industrial and Applied Mathematics},
  year      = {2018},
  series    = {MOS-SIAM Series on Optimization},
  doi       = {10.1137/1.9781611975444},
  isbn      = {978-1-61197-543-7}
}

@InProceedings{pmlr-v258-tokmak25aPACSBO,
  title = 	 {Safe exploration in reproducing kernel {Hilbert} spaces},
  author =       {Tokmak, Abdullah and Krishnan, Kiran G. and Sch{\"o}n, Thomas B. and Baumann, Dominik},
  booktitle = 	 {Proceedings of The 28th International Conference on Artificial Intelligence and Statistics},
  pages = 	 {784--792},
  year = 	 {2025},
  editor = 	 {Li, Yingzhen and Mandt, Stephan and Agrawal, Shipra and Khan, Emtiyaz},
  volume = 	 {258},
  series = 	 {Proceedings of Machine Learning Research},
  month = 	 {03--05 May},
  publisher =    {PMLR},
  url = 	 {https://proceedings.mlr.press/v258/tokmak25a.html}
}

@INPROCEEDINGS{PVKO_Parametric_Koopman,
  author={Lee, Changyu and Park, Kiyong and Kim, Jinwhan},
  booktitle={2023 62nd IEEE Conference on Decision and Control (CDC)}, 
  title={Parameter-Varying {Koopman} Operator for Nonlinear System Modeling and Control}, 
  year={2023},
  volume={},
  number={},
  pages={3700-3705},
  doi={10.1109/CDC49753.2023.10384235}}

@INPROCEEDINGS{BlischkeHespanha2023,
  author={Blischke, Madeline and Hespanha, Jo{\~a}o P.},
  booktitle={2023 62nd IEEE Conference on Decision and Control (CDC)}, 
  title={Learning Switched {Koopman} Models for Control of Entity-Based Systems}, 
  year={2023},
  volume={},
  number={},
  pages={6006-6013},
  doi={10.1109/CDC49753.2023.10384106}}

@article{rugh2000gain,
title = {Research on gain scheduling},
journal = {Automatica},
volume = {36},
number = {10},
pages = {1401-1425},
year = {2000},
issn = {0005-1098},
doi = {https://doi.org/10.1016/S0005-1098(00)00058-3},
url = {https://www.sciencedirect.com/science/article/pii/S0005109800000583},
author = {Wilson J. Rugh and Jeff S. Shamma}
}

@techreport{kraft1988slsqp,
  author      = {Dieter Kraft},
  title       = {A Software Package for Sequential Quadratic Programming},
  institution = {DFVLR, Institut f{\"u}r Dynamik der Flugsysteme},
  address     = {Oberpfaffenhofen, Germany},
  number      = {DFVLR-FB 88-28},
  year        = {1988}
}

\end{document}